\documentclass[aip,jcp,floatfix,10pt,groupedaddress,preprint]{revtex4-1}

\pdfoutput=1
\usepackage{siunitx}
\usepackage{tabularx, colortbl, xcolor, multirow}
\usepackage{amssymb, amsmath, dsfont, amsfonts, amsthm, amscd, mathtools}
\usepackage{graphicx, subfigure}
\usepackage{bm}

\usepackage{braket}

\begin{document}
\title{Interfacial charge rearrangement and intermolecular interactions: \\ Density-functional theory study of free-base porphine adsorbed on Ag(111) and Cu(111)}

\author{Moritz M\"uller}
\email[]{current address: CIC nanoGUNE Consolider, Tolosa Hiribidea 76, E-20018 Donostia - San Sebasti\'{a}n, Spain}
\affiliation{Department Chemie, Technische Universit{\"a}t M{\"u}nchen, \mbox{D-85747 Garching, Germany}}
\author{Katharina Diller$^*$}
\email[]{Author to whom correspondence should be addressed. Electronic mail: katharina.diller@tum.de}
\email[]{current address: Institute for Condensed Matter Physics, \'{E}cole polytechnique f\'{e}d\'{e}rale de Lausanne, CH-1015 Lausanne, Switzerland}
\affiliation{Department Chemie, Technische Universit{\"a}t M{\"u}nchen, \mbox{D-85747 Garching, Germany}}
\author{Reinhard J. Maurer}
\email[]{current address: Department of Chemistry, Yale University, New Haven CT 06520, USA}
\affiliation{Department Chemie, Technische Universit{\"a}t M{\"u}nchen, \mbox{D-85747 Garching, Germany}}
\author{Karsten Reuter}
\affiliation{Department Chemie, Technische Universit{\"a}t M{\"u}nchen, \mbox{D-85747 Garching, Germany}}

\begin{abstract}
We employ dispersion-corrected density-functional theory to study the adsorption of tetrapyrrole 2H-porphine (2H-P) at Cu(111) and Ag(111). Various contributions to adsorbate-substrate and adsorbate-adsorbate interactions are systematically extracted to analyze the self-assembly behavior of this basic building block to porphyrin-based metal-organic nanostructures. This analysis reveals a surprising importance of substrate-mediated van der Waals interactions between \mbox{2H-P} molecules, in contrast to negligible direct dispersive interactions. The resulting net repulsive interactions rationalize the experimentally observed tendency for single molecule adsorption. 
\end{abstract}

\maketitle 

\section{Introduction}
\label{intro}

The growing demand for further miniaturization of electronic components requires the exploration of new routes towards nanoscale devices with sizes below what can be reached with currently available top-down procedures such as photo-lithography. Possible solutions to approach smaller length scales comprise bottom-up techniques employing molecular building blocks that assemble into devices. Due to their large variety and tunable functionality organic molecules are often chosen as such basic building blocks. The ultimate goal is a rational design of molecular assembly by steering individual molecular interactions via material composition and molecular functionalization.

\begin{figure}
\includegraphics[width=0.5\columnwidth]{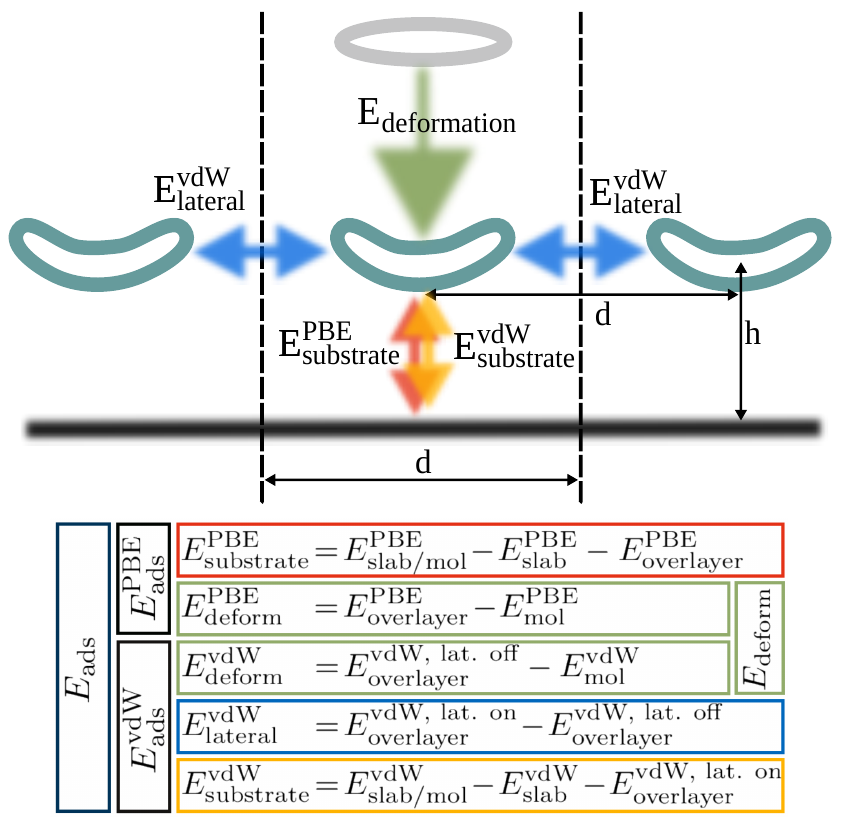}
\caption{Schematic depiction of different energy terms contributing to the adsorption energy. During the adsorption process the energy of the substrate-adsorbate system is modified due to the formation of the overlayer structure ($E_{\textrm{deform}}$), lateral van~der~Waals interactions ($E_{\textrm{lateral}}^{\textrm{vdW}}$), and substrate-induced energy contributions. The latter can be divided in a vdW part and all other contributions, in this context denoted as $E_{\textrm{substrate}}^{\textrm{PBE}}$. The table below compiles the defining equations of these contributions and other relevant such quantities (see text).}
\label{fig:forces}
\end{figure}

Surface self-assembly can be experimentally studied with scanning tunneling microscopy (STM), which allows a direct visualization in real space.~\cite{Binnig1982} Ultra-high vacuum conditions enable the controlled study of adsorbate-substrate interactions at a molecular level. However, often enough successful assembly can only be achieved by trial-and-error studies, simply exploring combinations of substrates, molecules and their functionalization. For a targeted design of devices we instead need a detailed understanding of all interaction components, such as the covalent bonding, penalties due to sterical hindrance, van der Waals contributions, or electric dipole formation due to e.g.\ charge transfer. These interactions can be classified in categories, such as molecule-adsorbate or adsorbate-adsorbate interactions (as shown in Fig.~\ref{fig:forces}). Varying the substrate or the functionalization within a group of molecules then aims at tuning different interactions.

\begin{figure}
\includegraphics[width=0.2\columnwidth]{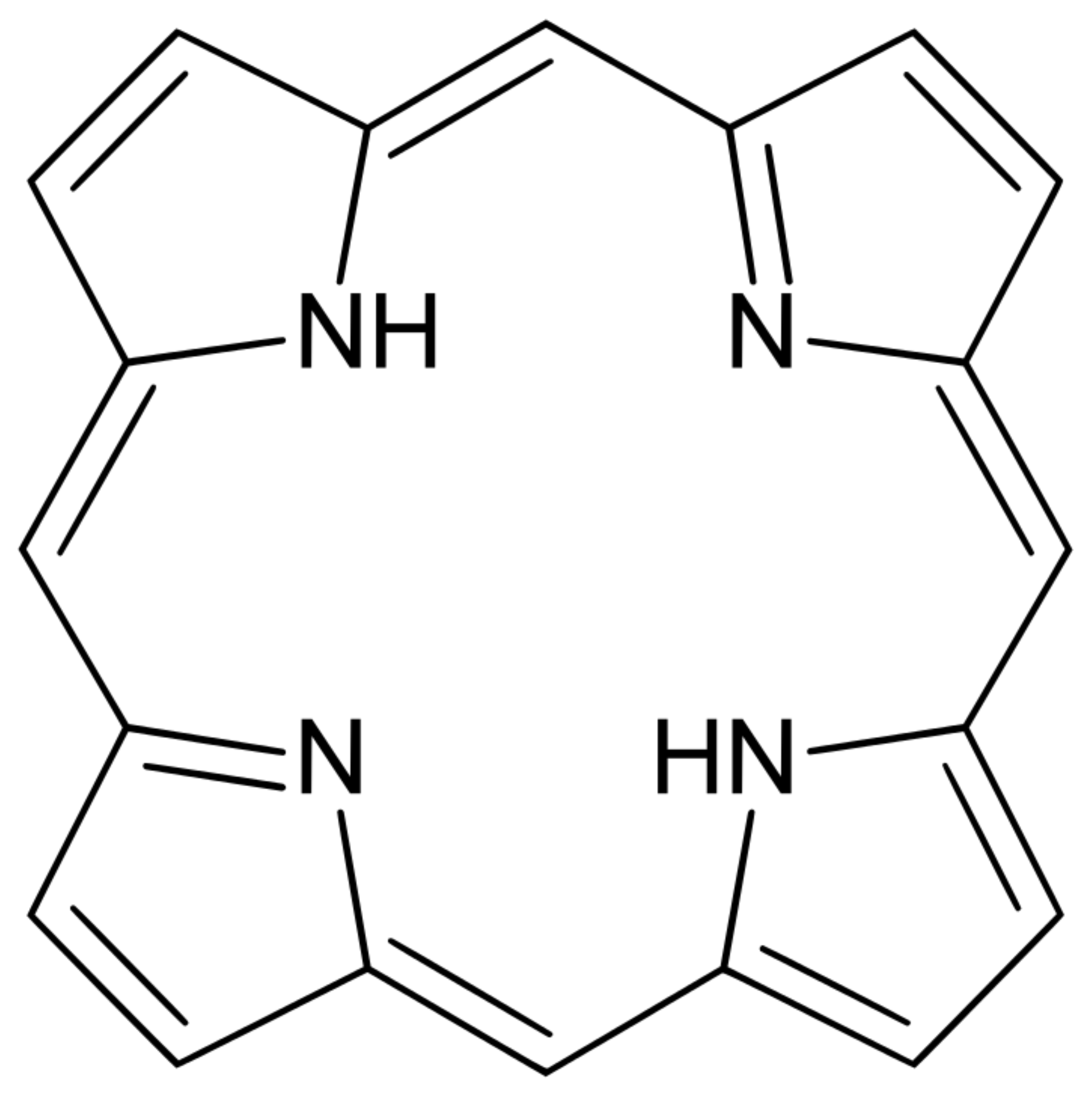}
\caption{\label{fig:2h-p} Structural formula of free-base porphine (2H-P)}
\end{figure}

A well-studied class of organic molecules in this context are porphyrins, which combine interesting electronic properties with a high structural variability. A plethora of investigations considers their technological potential, \emph{e.g.}, in molecular memory devices~\cite{Li2004,Liu2003}, photovoltaics~\cite{Campbell2004,Li2013a}, gas sensors \cite{Rakow2000}, light emission~\cite{Baldo1998}, and catalysis~\cite{Su2010}. All porphyrins share the same molecular unit, namely free-base porphine (\mbox{2H-P}, C$_{\textrm{20}}$H$_{\textrm{14}}$N$_{\textrm{4}}$, cf.\ Fig.~\ref{fig:2h-p}), which is the simplest porphyrin comprising four pyrrol rings linked by methine bridges. To functionalize the molecule a metal center can be included and/or a wide range of substituents can be attached. This strongly affects the self-assembly properties of the system. On Cu(111), for example, a range of different structures is observed, from molecular chains or nanoporous networks~\cite{Wintjes2008} over close-packed islands~\cite{Xiao2012} to individual molecules.~\cite{Rojas2010,Diller2013} The role of the substrate becomes for instance evident for free-base tetraphenyl-porphyrin (\mbox{2H-TPP}), which does not self-assemble on Cu(111), but was reported to do so on Ag(111).~\cite{Rojas2010}

Given its importance as basic porphyrin unit, it is surprising that the interaction of \mbox{2H-P} with surfaces only recently became a point of interest, while its crystal structure~\cite{Webb1965} and electronic structure~\cite{Orti1989,Blase2011a,Dabo2013,Zhang2005} are already well studied. STM experiments show that both on the Cu(110)~\cite{Haq2011} and the Cu(111)~\cite{Diller2013} surface the \mbox{2H-P} molecules remain isolated. Intriguingly, also on the more inert Ag(111) surface, where substituted porphyrins often assemble into islands~\cite{Rojas2010,Iacovita2012}, the \mbox{2H-P} units avoid each other.~\cite{Bischoff2013a} One possible explanation for this behavior are repulsive intermolecular forces due to interfacial charge redistribution (found experimentally for both Ag(111)~\cite{Bischoff2013a} and Cu(111)~\cite{Diller2013}) which leads to repulsive dipoles. In agreement with this assumption \mbox{2H-P} molecules deposited on Ag(111) were found to form islands in the second layer~\cite{Bischoff2013a}, emphasizing the role of the immediate molecule-substrate interactions. The experimental results as such, however, cannot completely describe the system. An often postulated charge transfer between the adsorbates and the metal support (especially when as weak as likely for \mbox{2H-P}/Ag(111)) does not necessarily lead to isolated adsorbate species. This has been demonstrated for the case of \mbox{2H-TPP} on Ag(111) and Cu(111), where the charge rearrangement competes with many other attractive, but also repulsive effects.~\cite{Rojas2010} A detailed analysis of the different interaction contributions depicted in Fig.~\ref{fig:forces} for 2H-P is therefore at place. 

In this work we partition the adsorption energy into different physically-motivated components and quantify the different contributions that govern the self-assembly behavior of an adsorbate on a metal surface. We use this approach for the prototypical test case \mbox{2H-P} adsorbed on the (111)-facets of silver and copper by means of dispersion-corrected density functional theory. For both substrates, an initial analysis of the low-coverage limit allows to disentangle covalent and dispersion contributions, as well as the role of charge-transfer. In the study of further contributions shown in Fig.~\ref{fig:forces} we then take advantage of the fact that we can selectively switch on and off vertical and horizontal van der Waals interactions, simulate different coverages by varying the surface unit-cell size, or remove the substrate. This allows to particularly discuss the dependence of the individual energy contributions as functions of surface coverage. In agreement with experiment we find that there is no strong energetic driving force for island formation on either surface and arrive at a rationalization of the missing self-assembly notion in the case of \mbox{2H-P} on Ag(111). Finally, we discuss molecular functionalization and substrate modification as potential routes for a rational interaction tuning.

\section{Computational Details and Methods}
\label{sec:methods}

All calculations in this work were carried out with the periodic plane-wave Density-Functional Theory (DFT) code CASTEP 6.0.1,~\cite{Segall2002,Payne1992,Clark2005} along with standard library ultrasoft pseudopotentials~\cite{Vanderbilt1990}. Short-range electronic exchange and correlation are treated using the semi-local PBE functional~\cite{Perdew1996}. Long range van~der~Waals (vdW) forces and the collective many-body response of the metallic substrate~\cite{Ruiz2012} are considered through effective pairwise-additive dispersion correction scheme vdW$^\mathrm{surf}$~\cite{Tkatchenko2009, Ruiz2012}. The resulting PBE+vdW$^\mathrm{surf}$ scheme was found to yield good results for induced dipoles, as well as the density of states of planar aromatic adsorbates on metal surfaces.~\cite{Hofmann2013}

The Atomic Simulation Environment (ASE)~\cite{Bahn2002} was used to set up the simulation cells (vacuum $>$~\SI{20}{\angstrom}). Each slab was constructed as a perfect (111)-fcc surface with PBE-optimized lattice constants of \SI{4.14}{\angstrom} for bulk Ag and \SI{3.63}{\angstrom} for bulk Cu. Four metal layers, sufficient to describe the ABCABC stacking scheme and shown to be sufficient to describe the system~\cite{Hanke2011,Haq2011,Dyer2011}, were employed in the calculations.

Computational settings for energy cutoff, k-point sampling, and vacuum between the slabs converge the adsorption energies to within \SI{+-20}{\meV}. The latter energies are defined as
\begin{equation}
E_{\textrm{ads}} = E_{\textrm{slab/2H-P}} - \left(E_{\textrm{slab}} + E_{\textrm{2H-P}}\right) \quad,
\label{eq:E_ads}
\end{equation}
with $E_{\textrm{slab}}$ the total energy of the clean slab, $E_{\textrm{2H-P}}$ the total energy of the isolated gas-phase molecule and $E_{\textrm{slab/2H-P}}$ that of the combined system. For geometry optimizations we use an energy cutoff of \SI{450}{\eV}, and a \mbox{2 $\times$ 2 $\times$ 1} Monkhorst-Pack grid \cite{Monkhorst1976}. Final single-point calculations employed the same energy cutoff, but a more densely sampled (\mbox{4 $\times$ 4 $\times$ 1}) k-point grid, in order to resolve the adsorption energy with the desired accuracy. As indicated in sections \ref{sec:substrate_adsorbate_interactions} and \ref{sec:interadsorbate_interactions}, the size of the surface unit-cell was varied between \num{5x5}, \num{6x6} and \num{7x7} times the primitive hexagonal surface-unit cell to simulate different molecular coverages. The different contributions to the total adsorption energy for these three coverages were defined as shown in Fig.\ \ref{fig:forces} (see SI for a more detailed description).  

The geometry optimizations of the adsorption structures were performed for frozen substrates using a delocalized internal coordinates optimizer\cite{Andzelm2001} and an ionic force tolerance of \SI{25}{\meV\per\angstrom} per atom, an energy tolerance of \SI{2e-5}{\eV}. The electronic structure was  converged with an energy tolerance of \SI{e-8}{\eV}. A Gaussian type electronic smearing procedure with a width of \SI{0.1}{\eV} was employed.

Charge partitioning was performed with the Bader,\cite{Bader1991, Bader1994} Mulliken,\cite{Mulliken1955c,Mulliken1955b,Mulliken1955a,Mulliken1955,Sanchez-Portal1995} and Hirshfeld\cite{Hirshfeld1977} methods as implemented in CASTEP and provided by the BADER tool\cite{Henkelman2006,Sanville2007,Tang2009}. In addition, we also employed a projection of the total density of states (DOS) onto the molecular orbitals (MO-PDOS) of the free-standing overlayer, i.e., a layer of \mbox{2H-P} molecules in their adsorbed configuration but without metal support. The projected density of states $\rho_j(E)$ with respect to reference gas-phase MO~$\phi_i$ of the free standing overlayer is given by 
\begin{equation}
 \rho_j(E)=\sum_i|\langle \psi_j|\phi_i\rangle|^2\delta(E-\varepsilon_i),
 \label{eq:pdos}
\end{equation}
where $\psi_j$ are the calculated Kohn-Sham orbitals and $\varepsilon_i$ the corresponding energy eigenvalues. Integration up to the Fermi level $E_F$ yields the occupation of the molecular orbital $\phi_i$. The charge on an adsorbed \mbox{2H-P} molecule can then be determined by summing up all the occupations of the MOs. Higher-lying unoccupied orbitals are not described well due to the exponential decay of the semi-local xc-potential, thus only contributions up to the second lowest unoccupied MO (LUMO+2) were considered for the quantitative analysis. We expect this to result in only a small error for the quantification of charge transfers, because already the LUMO+2 lies almost entirely above the Fermi level (Fig.~\ref{fig:DOS}).

\section{Results and Discussion}
\label{results}

We first investigate the low-coverage limit of \mbox{2H-P} on Ag(111) and Cu(111) to arrive at a first quantification of covalent and dispersive bonding contributions, the adsorption induced charge re-arrangement, as well as charge-transfer. Subsequently we analyze the individual contributions depicted in Fig.~\ref{fig:forces} as a function of surface coverage.

\subsection{Adsorbate-Substrate Interactions}

\label{sec:substrate_adsorbate_interactions}
\begin{figure}
\begin{center}
\label{fig:bridge30_cu}
\includegraphics[width=0.48\columnwidth]{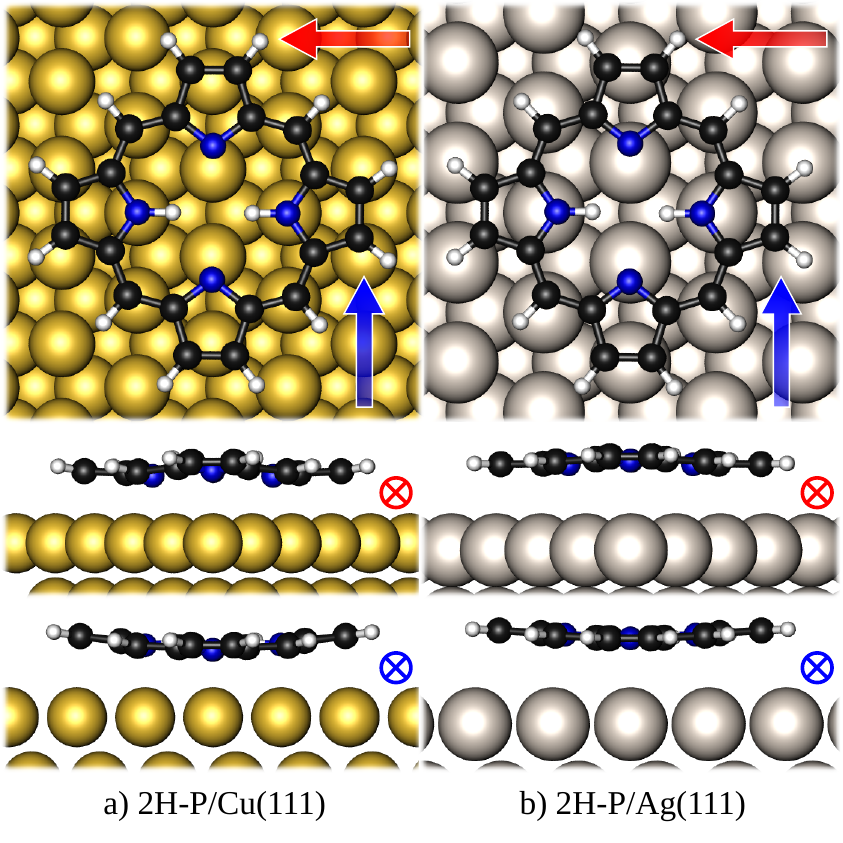}
\end{center}
\caption{\label{fig:geometries}
Calculated adsorption geometries of porphine adsorbed at the bridge sites of \mbox{Cu(111)} (left) and \mbox{Ag(111)} (right). View from the top (top panels) and from the sides; viewing directions are indicated by the red horizontal (middle panels) and the blue vertical arrow (bottom panels). Adsorption at the more reactive copper surface leads to lower adsorption heights and a stronger deformation of the adsorbate.}
\end{figure}

\begin{figure}
\includegraphics[width=0.48\columnwidth]{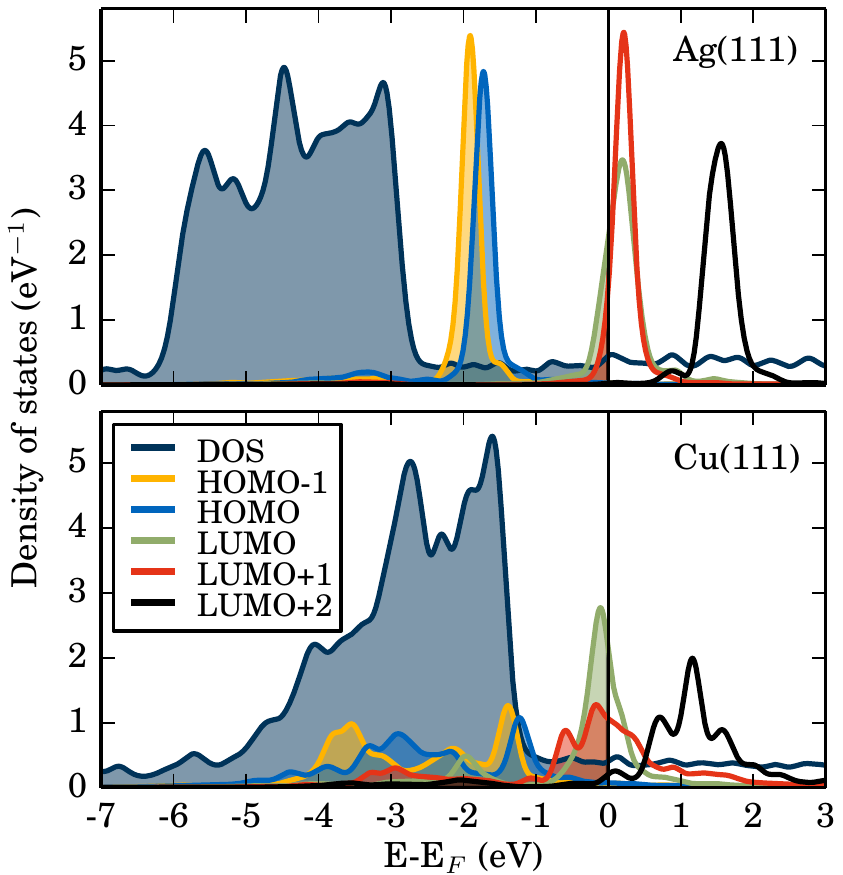}
\caption{\label{fig:DOS} Total density of states (DOS, blue, downscaled by a factor of 130) and molecular-orbital projected density of states (MO-PDOS) of \mbox{2H-P} on \mbox{Ag(111)} (top) and \mbox{Cu(111)} (bottom), indicating the changes of the frontier orbitals due to adsorption on the respective surfaces. Energies are given with respect to the Fermi level $E_{\rm F}$. On \mbox{Ag(111)} the peak shape of the free molecules is nearly retained, while on Cu(111) the peak splitting and broadening point to a stronger hybridization. Down-shifting of the LUMOs below $E_{\rm F}$ (shaded areas) indicate a charge transfer to the molecule.}
\end{figure}

To analyze the interaction between a single \mbox{2H-P} molecule and the Ag(111) or Cu(111) surface we place one \mbox{2H-P} in a (6 $\times$ 6) surface unit-cell, corresponding to a distance of 15.4\,{\AA} between the molecule and its periodic images. To ensure that this setup represents the low-coverage limit the van der Waals interactions between the adsorbates in neighboring cells are switched off. Considering all possible high-symmetry adsorbate positions (centered at top, bridge, fcc and hcp hollow sites) and orientations, the most stable adsorbate geometry on both substrates is found to be centered above the bridge site as displayed in Fig.~\ref{fig:geometries}. This geometry results in adsorption energies of \SI{-2.60}{\eV} on Ag(111) and \SI{-3.15}{\eV} on Cu(111). These findings are consistent with the experimental analysis of the adsorption patterns at submonolayer coverages, which equally suggested bridge positions as preferred adsorption sites on Ag(111).~\cite{Bischoff2013a} The stronger adsorption on copper is accompanied by lower adsorption heights (\SI{2.40}{\angstrom} on Cu(111) and \SI{2.89}{\angstrom} on Ag(111), center of mass) and a slightly stronger deformation of the adsorbate (cf.\ side views in Fig.~\ref{fig:geometries}). The maximum tilt angles of the pyrrole planes (\SI{7}{\degree} on Cu, \SI{4}{\degree} on Ag) again agree well with results from angle-resolved X-ray absorption spectroscopy, which predicted a flat and planar adsorption (with a slight deformation on copper) of the molecules at submonolayer coverages.~\cite{Bischoff2013a,Diller2013,Diller2014} The adsorption energy on Cu(111) is \SI{0.32}{\eV} lower than the one determined previously by vdW-DF calculations for \mbox{2H-P/Cu(110)}.~\cite{Dyer2011} While this comparison is blurred by the different types of vdW-corrections employed in the two studies, it seems plausible that adsorption is indeed slightly weaker on the closed-packed Cu(111) surface. In general, the absence of substituents allows the macrocycle to adsorb closer to the surface compared to for example \mbox{2H-TPP} on Ag(111) and Cu(111), where the nitrogens are found at a distance of $>$~\SI{3}{\AA} from the substrate.~\cite{Rojas2012}

Fig.~\ref{fig:DOS} depicts the total density of states (DOS) and the molecular orbital projected density of states (MO-PDOS) of the frontier orbitals of \mbox{2H-P} adsorbed on Ag(111) (top) and Cu(111) (bottom), which shows how the orbitals of the gas-phase molecule change upon adsorption. For isolated \mbox{2H-P} molecules in the gas-phase (not shown) the LUMO and the LUMO+1 are nearly degenerate, as is typical for free-base porphyrins.\cite{Zhang2005} The interaction with the silver surface partially lifts this degeneracy of LUMO (red) and LUMO+1 (green) and causes both MOs to become partially occupied (Fig.~\ref{fig:DOS}, top). In general, however, the MO-PDOS peaks mainly retain the same character as in the free molecules, pointing to a relatively weak chemical interaction with the substrate. This is not the case for \mbox{2H-P/Cu(111)} (Fig.~\ref{fig:DOS}, bottom), where the states are strongly shifted, split, and broadened, as is characteristic for chemisorbed systems. The filling of the LUMO and the LUMO+1 is correspondingly much stronger than for the adsorption at Ag(111), in agreement with the experimental predictions.~\cite{Diller2013,Bischoff2013a} Apart from the actual shift, the most striking effect of the substrate is the change of the peak shapes: While the single, molecular-type states are still discernible on Ag(111), they nearly vanish for Cu(111) and are replaced by band-like structures caused by peak splitting and broadening as is typical for hybridized systems. This suggests a different nature of the bond at the two substrates: Porphine is rather physisorbed on silver, while it is closer to chemisorbed on Cu(111). 

\renewcommand{\arraystretch}{1.2}
\begin{table}[t]
 \centering
 \caption{Energy contributions (cf.\ definitions in Fig.~\ref{fig:forces}) to the adsorption energy for \mbox{2H-P} on \num{6x6} Cu(111) and Ag(111) slabs (also cf.\ section \ref{sec:interadsorbate_interactions} and Fig.\ \ref{fig:coverage}).}
\begin{tabular}{c|ccccc}
\toprule
 & $E_{\textrm{ads}}$ & $E_{\textrm{deform}}$ &  $E_{\textrm{lat.}}^{\textrm{vdW}}$ & $E_{\textrm{substr.}}^{\textrm{vdW}}$ & $E_{\textrm{substr.}}^{\textrm{PBE}}$ \\ \hline
\ Ag(111)\ & \ -2.60~eV\ & \ 0.08~eV\ & \ 0.00~eV\  & \ -2.99~eV\  & \ 0.30~eV\ \\ 
\ Cu(111)\ & \ -3.15~eV\ & \ 0.67~eV\ & \ 0.00~eV\  & \ -3.85~eV\  & \ 0.03~eV\ \\ \hline 
\end{tabular}
\label{tab:energies_single}
\end{table}
\renewcommand{\arraystretch}{1}

This interpretation is further supported by the differing adsorption energy of \mbox{2H-P} on silver (-2.60~eV) and copper (-3.15~eV). Disentangling the energy contributions as defined in Fig.~\ref{fig:forces} and listed in Table~\ref{tab:energies_single} (cf.\ also Fig.\ \ref{fig:coverage}) we elucidate the main causes for this difference. From the definition in Eq.~\ref{eq:E_ads} it thereby follows that energy terms with $E<0$ are attractive, $E>0$ are repulsive. On both substrates, the only attractive (and at the same time also dominating) contribution is the van der Waals interaction between molecule and substrate, while $E_{\textrm{deform}}$ and $E_{\textrm{substr.}}^{\textrm{PBE}}$ contribute repulsively. Interesting are the differences between Ag and Cu: While on copper the deformation energy is the most repulsive term (consistent with the more pronounced deformation as described above) and $E_{\textrm{substr.}}^{\textrm{PBE}}$ nearly vanishes, the situation is reversed on silver. Taking into account the level of hybridization this suggests that on copper also ``attractive`` covalent contributions are present. This means that the stronger adsorption on copper is not only the result of a stronger van der Waals interaction, but also of ``less repulsive`` short-range covalent contributions as picked up by the semi-local DFT functional.

\begin{figure}
\includegraphics[width=0.48\columnwidth]{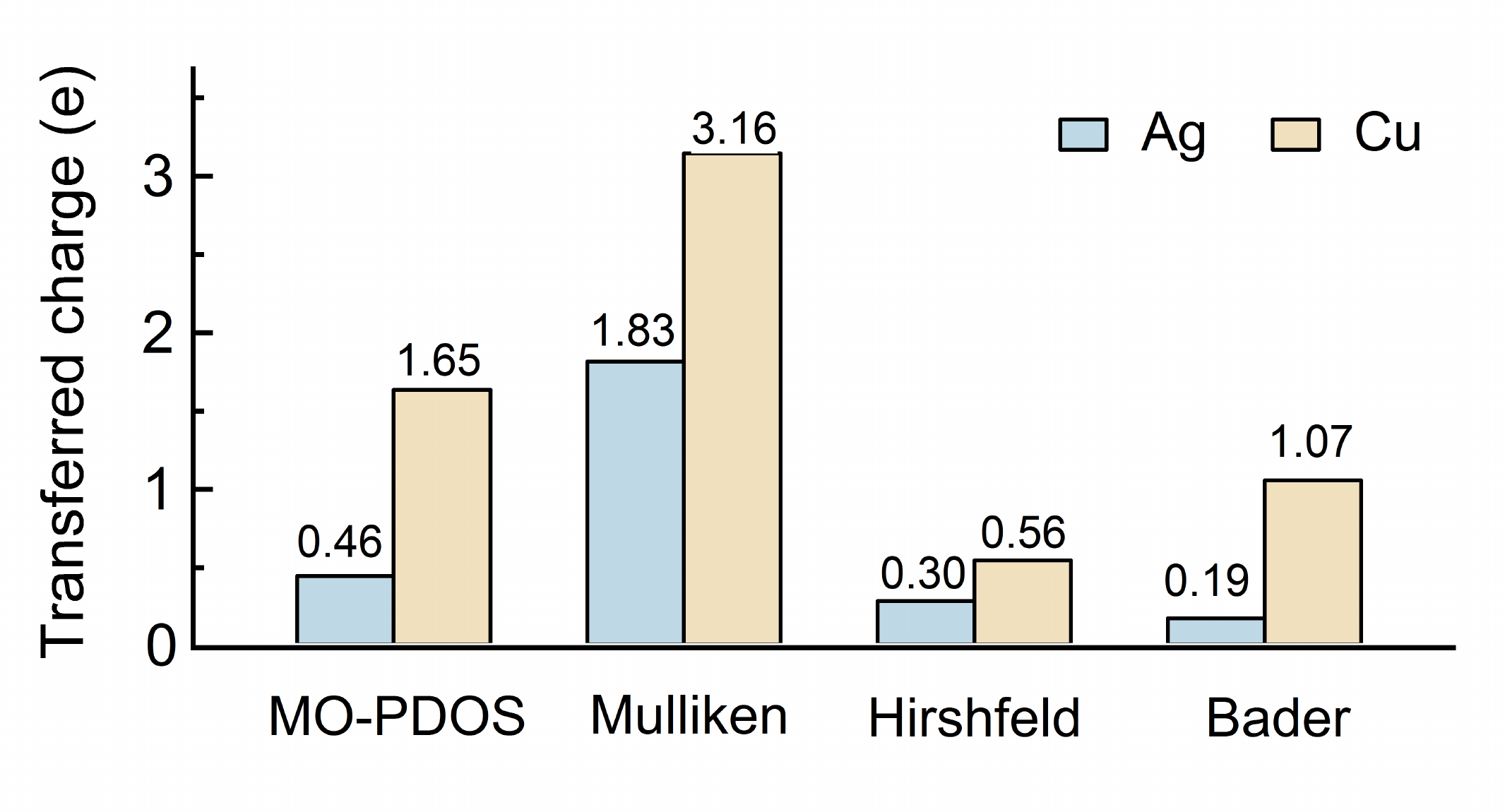}
\caption{\label{fig:el_transfer} Quantitative comparison of the net electron transfer (in units of elementary charge \textit{e}) from the Ag(111) (blue) and the Cu(111) (orange) surfaces to the \mbox{2H-P} molecule determined with different methods. See also Table S1 for coverage dependent values.}
\end{figure}

As mentioned initially, interfacial charge redistribution leading to repulsive dipoles between adsorbed molecules has been put forward as one possible explanation for the lacking self-assembly tendency of porphins observed on both Ag(111)~\cite{Bischoff2013a} and Cu(111)~\cite{Diller2013}. Scrutinizing this hypothesis, we compare the results from different partial charge partitioning schemes, namely Mulliken,\cite{Mulliken1955c,Mulliken1955b,Mulliken1955a,Mulliken1955,Sanchez-Portal1995} Hirshfeld\cite{Hirshfeld1977} and Bader.\cite{Bader1991, Bader1994} 
It is well known that absolute partial charge values can vary significantly between different partitioning schemes~\cite{Wiberg1993,Leenaerts2008,Jacquemin2012}. For the present purposes we therefore only focus on the relative trends between both substrates. The calculated overall net charge on the molecule after adsorption (cf. Fig.~\ref{fig:el_transfer}) confirms the interpretation derived from the MO-PDOS calculation: On both substrates charge is indeed transferred towards the porphine molecules and to a substantially higher extent on Cu(111) than on Ag(111). Employing different charge partitioning methods allows furthermore for a quantitative comparison with literature values for similar porphyrin-substrate systems obtained with one or the other of these schemes (Table \ref{tab:CT_comp}). This comparison yields two main results for \mbox{2H-P} on Ag(111) and Cu(111): (i) the values for both substrates are comparable with those of substituted compounds, and even metalloporphyrins, and (ii) the general trend between the partitioning schemes (i.e., Mulliken yields the highest values) is in agreement with that in Table \ref{tab:CT_comp} and what is generally known for the different partitioning schemes.\cite{Wiberg1993,Leenaerts2008,Jacquemin2012} 

\begin{figure}
\includegraphics[width=0.42\columnwidth]{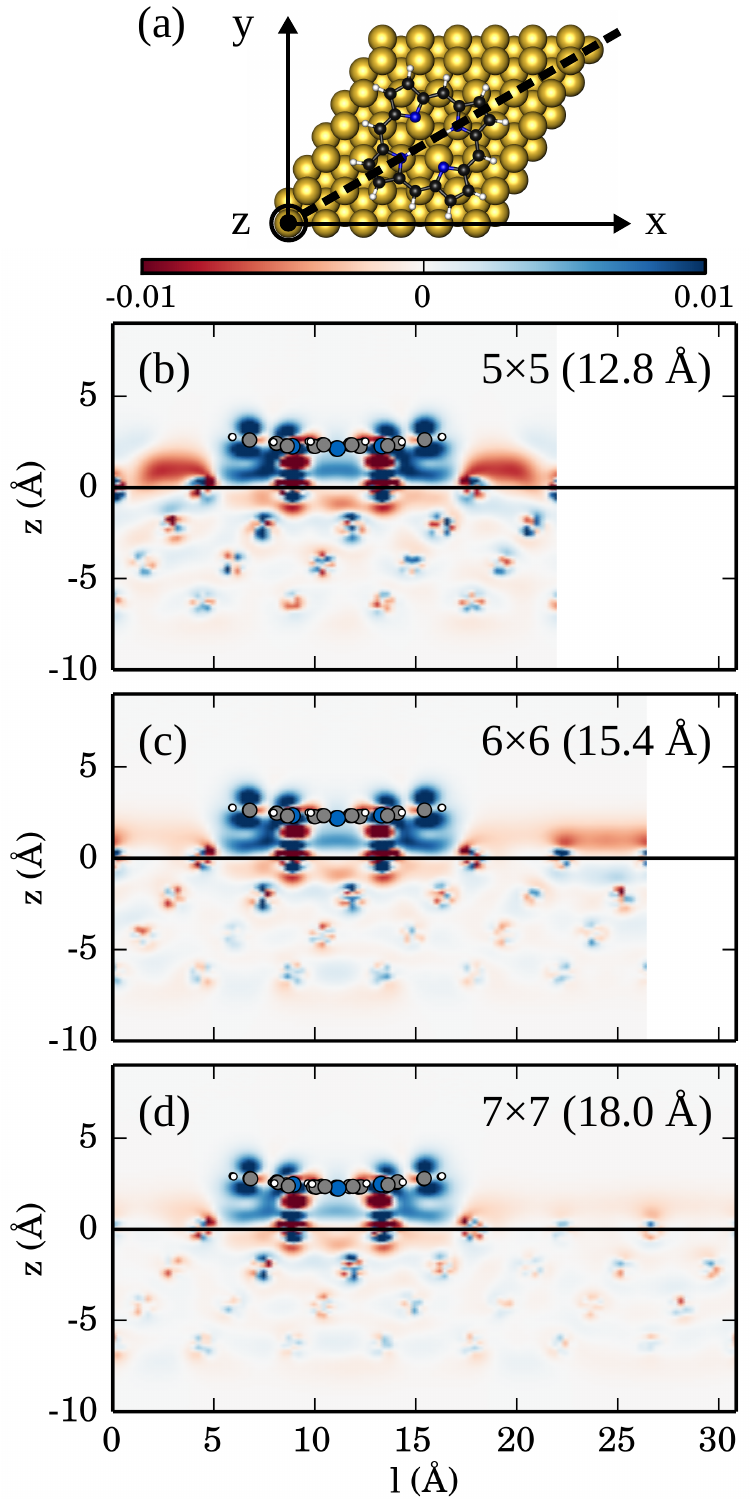}
\caption{\label{fig:sections} Charge density difference of \mbox{2H-P} after adsorption onto Cu(111) for different coverages. (b) Cuts through the unit cell along the black line indicated in (a). The centers of the top copper atoms are placed at z = 0 (z $<$ 0: location inside the substrate). Regions of electron accumulation (depletion) are depicted in red (blue), values are given in $e \textrm{\AA}^{-3}$. Values larger than $\pm$ 0.01 are mapped onto $\pm$ 0.01. With increasing coverage the accumulation of charge between molecules increases substantially.}
\end{figure}

\renewcommand{\arraystretch}{1.0}
\begin{table}[b]
 \centering
 \caption{Comparison of our calculated charge transfer values for free-base porphine with published literature data for substituted porphyrins.}
\begin{tabular}{c|ccc}
\toprule
method & \ Ag(111)\ & \ Cu(111)\ &  literature \\ \hline
\multirow{2}*{Mulliken} & \multirow{2}*{1.83~\textit{e}} & \multirow{2}*{3.16~\textit{e}} & \  2H-TPP/Cu(111): 1.69~\textit{e}\cite{Rojas2012} \\ 
& & & \ 2H-TPP/Ag(111): 1.45~\textit{e}\cite{Rojas2012} \\ [2.5mm]

\multirow{4}*{Bader} & \multirow{4}*{0.19~\textit{e}} & \multirow{4}*{1.07~\textit{e}} & \quad 2H-P/Cu(110): 0.9~\textit{e}\cite{Dyer2011} \quad\\ 
& & & Co-TPP/Ag(111): 0.37~\textit{e}\cite{Hieringer2011} \\
& & & Fe-TPP/Ag(111): 0.22~\textit{e}\cite{Hieringer2011} \\
& & & Cu-TBPP/Si(111)-B: 0.17~\textit{e}\cite{Boukari2012} \\ [2.5mm]

\ \multirow{2}*{Hirshfeld}\quad & \multirow{2}*{0.30~\textit{e}} & \multirow{2}*{0.56~\textit{e}} &  \ 2H-TPP/Ag(111): 0.46~\textit{e}\cite{Rojas2012} \\
& & & \ 2H-TPP/Cu(111): 0.89~\textit{e}\cite{Rojas2012} \\ \hline 
\end{tabular}
\label{tab:CT_comp}
\end{table}
\renewcommand{\arraystretch}{1}

\begin{figure}
\includegraphics[width=0.48\columnwidth]{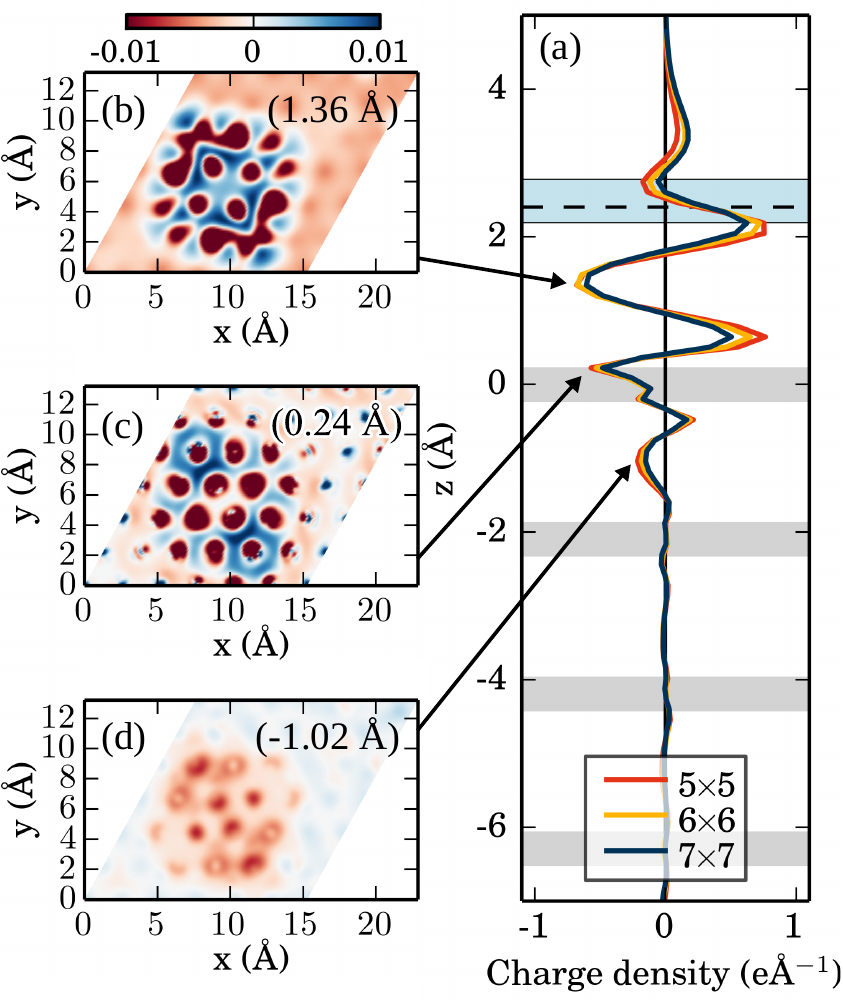}
\caption{\label{fig:sections_vert} Charge density difference (a) after integrating in x and y and (b) for horizontal cuts along selected peaks of (a). Negative values and red regions refer to an increase in electronic density. The four gray horizontal bars indicate the position of the atomic layers in the Cu slab. The dashed black horizontal line displays the center of mass of the adsorbed molecule and the region shows the minimum and maximum components of the atomic positions.}
\end{figure}

In principle, we could also use the partial charge analysis to determine where the excess charge is located. However, using the atomic charge location as given by the partial charges of the two different nitrogen species (N and NH) in porphine as probe, it becomes clear that one has to proceed with caution: Not only do the values vary quantitatively as could be expected from the net charge results presented above. Instead, even the qualitative interpretation changes for the different methods (Fig.~S1, see discussion in section 1 of the Supporting Information).\cite{SuppInf} The reason becomes clear when visualizing the charge distribution as density-difference plots illustrating the change in electron density after the adsorption of the molecule. For \mbox{2H-P} on Cu(111) in the \num{6x6} surface unit-cell we first consider the vertical density difference along the diagonal of the unit cell (Fig.~\ref{fig:sections}a, black line). The corresponding graph (Fig.~\ref{fig:sections}b, middle panel) shows that the net charge transfer discussed above does not properly reflect the complicated charge re-distribution upon adsorption of the molecule: There are regions with charge accumulation (red), as well as depletion (blue). The shape of the distribution is thereby typical for the so-called ``pillow effect`` where Pauli repulsion causes an electron accumulation around the edges of the molecule, a charge depletion directly below the molecule and the push-back of electron charge into the substrate (i.e., below the horizontal black line in Fig.~\ref{fig:sections}b at $z=0$).~\cite{Rojas2012} This is even more evident from the plot in Fig.~\ref{fig:sections_vert}a where the density was integrated in x and y and is then only shown as a function of the distance from the surface $z$. 

The shape of this plot, as well as the horizontal cuts (Fig.~\ref{fig:sections_vert}b-d) resemble those of 2H-P/Cu(110)~\cite{Dyer2011} and illustrate well that ``net charge transfer between surface and molecule`` does not fully describe the complexity of the charge redistribution upon surface adsorption. The strong peak halfway between the molecule and the surface indicates a maximum of charge-density accumulation. The corresponding horizontal cut in Fig.~\ref{fig:sections_vert}b closely resembles the shape of the combined LUMO and LUMO+1 of \mbox{2H-P} analogous to the case of 2H-P/Cu(110)~\cite{Dyer2011}. Fig.~\ref{fig:sections_vert}d shows how electrons are pushed back into the substrate, such that even between the first and second copper layer the imprint and shape of the porphine is still visible. Consistently the surface work function is reduced from 4.66 eV to 4.15~eV (Table \ref{tab:Workfunction}) upon adsorption of \mbox{2H-P}. This is typical for systems with a pronounced pillow effect as the electrons are pushed further away from the surface and the potential above the surface is correspondingly reduced.~\cite{Bagus2002}

Figs.~\ref{fig:sections} and \ref{fig:sections_vert} also illustrate the strong dependence of the partial charges on the employed method: The lateral and vertical fluctuations in the density difference are high, thus the way how charge is assigned to atoms in the molecule plays a big role. In principle it should also be possible to integrate the plot in Fig.~\ref{fig:sections_vert}a to obtain the net charge transfer towards the molecule. The crucial point is, however, how ''charge on the molecule``, or, more drastically put, how the ''molecule`` is defined in this picture. By changing this definition (e.g., taking atomic positions or the space halfway between the molecule and the substrate) largely varying numbers can be produced, so that using established partitioning schemes seems more consistent. We observe similar findings for \mbox{2H-P} on Ag(111), but to a weaker extent (cf. Supporting Information).\cite{SuppInf}

At this point it seems advisable to mention a general shortcoming of \emph{a-posteriori} dispersion corrected DFT+vdW or DFT-D treatments. Whereas the adsorption geometry and height are largely controlled by the vdW contribution, on the semi-local DFT side the molecule is pushed into a repulsive regime (due to the missing self-consistent coupling between vdW and DFT). A recently published implementation of a self-consistent vdW scheme~\cite{Ferri2015} has shown that this effect only minimally modifies geometries, but can significantly alter electronic structure in form of the work function or charge distribution.

Nevertheless, we can attempt to estimate this effect by analyzing the dependence of our results on the portion of covalent interaction between adsorbate and substrate. We can do this by reducing the covalent adsorbate-substrate interactions by adding a non-local functional that penalizes the LUMO orbital resulting in a higher LUMO energy and a reduced charge-transfer:
\begin{equation}
 \hat{V}_{c}=\frac{U}{2}|\phi_c\rangle \langle\phi_c|
\end{equation}
In the above penalty function $V_c$ we effectively apply a penalty $U$ on gas-phase molecular reference orbitals $\phi_c$ (see SI for more details).
With a penalty of 2 or 3~eV in the current formalism we can reduce charge transfer between adsorbate and substrate on copper to the same level as found in Ag (see Fig. \ref{fig:coverage} or Fig. S3). Although the corresponding DFT repulsion is now very strong, E$_{\mathrm{substr.}}^{\mathrm{vdW}}$ remains almost unchanged, suggesting no strong coupling between covalent and vdW contributions in this system. We can therefore assume that qualitative trends across substrates and coverages for adsorption on these comparably weakly interacting surfaces are unaffected by the \emph{a-posteriori} vdW treatment.

\subsection{Adsorbate-Adsorbate Interactions}
\label{sec:interadsorbate_interactions}

We proceed by explicitly studying dependence of all previously discussed energy components and binding characteristics as a function of coverage. We do this by systematically increasing the surface unit-cell size from a \mbox{5 $\times$ 5} unit cell, over the previously employed \mbox{6 $\times$ 6} unit cell to a \mbox{7 $\times$ 7} unit cell. In all cells we now account for lateral van der Waals interactions between unit cells (as opposed to the previous section), which correspondingly allows to directly compare to real coverage effects.

\begin{figure}
\includegraphics[width=0.48\columnwidth]{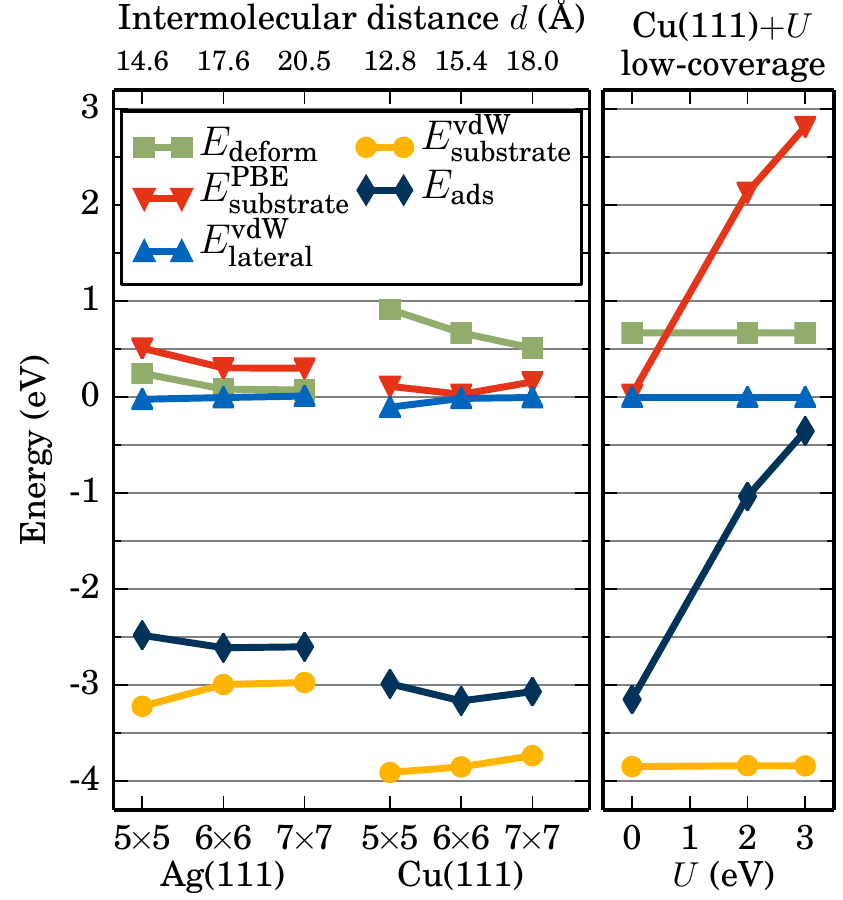}
\caption{\label{fig:coverage} Left panel: Calculated energy contributions to the total adsorption energy $E_{\textrm{ads}}$ under variation of the size of the surface unit-cell (i.e., the coverage) for \mbox{2H-P} on \mbox{Ag(111)} and \mbox{Cu(111)}. Lateral, intermolecular van der Waals interactions are negligible, while the van der Waals contribution between molecule and substrate dominate $E_{\textrm{ads}}$. The increase of $E_{\textrm{ads}}$ at higher coverages indicates that \mbox{2H-P} does not prefer to assemble into islands. Colors correspond to those in Fig.~\ref{fig:forces}. Right panel: Influence of the U-correction on the contributions to the adsorption energy of \mbox{2H-P} on Cu(111) in the low-coverage limit (\mbox{6 $\times$ 6} cell). The vdW-contributions are nearly unaffected.}
\end{figure}

The different energy contributions are displayed in Fig.~\ref{fig:coverage}. The overall tendency is the same for all coverages: The deformation energy and short-range DFT contribution are clearly repulsive ($E>0$) and lateral interactions are especially for silver nearly non-existent. The dominating attractive contributions are van der Waals interactions with the substrate which leads in all cases to stable total adsorption energies ($E<0$). Interesting in the context of self-assembly or island formation is the behavior at high coverages. On both substrates a similar trend can be observed in this respect: all van der Waals interactions become more attractive, whereas the deformation energy and DFT contributions become more repulsive for the smallest considered \mbox{5 $\times$ 5} cell. To understand this trend we compare the charge transfer (see Table I in SI) and density differences (Fig.~\ref{fig:sections} and \ref{fig:sections_vert}) on both substrates. With increasing coverage both the charge transfer and the pillow effect are increased, and consequently the work function is also further reduced (Table~\ref{tab:Workfunction}). Whereas work functions in Table III are approximated from the mid-point vacuum potential of an asymmetric slab, we do not expect the remaining deviations from dipole-corrected slabs to influence the overall observed trends. For both substrates the interplay of the laterally repulsive deformation energy and laterally attractive van der Waals contributions with the substrate results in very shallow minima (difference between \mbox{6 $\times$ 6} and \mbox{7 $\times$ 7}: 0.097~eV on Cu, 0.026~eV on Ag) in the total adsorption energy for the \mbox{6 $\times$ 6} unit cell (-2.61~eV for Ag(111) and -3.16~eV for Cu(111)). The strong variation of the $E_{\mathrm{substr.}}^{\mathrm{vdW}}$ is an interesting finding, as it means that the distance between adsorbates indirectly influences the bonding strength of the adsorbate with the substrate, thereby effectively yielding a \emph{via}-substrate or substrate-mediated interaction between adsorbates.

Inspecting the employed vdW$^\textrm{surf}$ scheme, the observed \emph{via}-substrate vdW-interaction can only be traced back to either changes in the adsorbate structure or changes in the electron density. The latter modify the C$_6$­ coefficients indirectly since those are functionals of the electronic density through a connection via Hirshfeld partitioning. We analyze our results for the second possibility across different coverages (\mbox{5 $\times$ 5} to \mbox{7 $\times$ 7} cell) by effectively switching off the dependency of the vdW scheme on the electronic density. This is achieved by setting the Hirshfeld volume of a bound atom equal to the reference value of corresponding free atom. The analysis suggests that density fluctuations are visible in the resulting energy contributions but play only a small role with respect to the formation of overlayers in our systems ($\Delta E_{\mathrm{substr.}}^{\mathrm{vdW}}$ of 0.03~eV for both substrates). We hence relate the effective substrate-mediated interactions to structural changes where the increase of adsorption energy at higher coverage follows a reduction of adsorption height ($\Delta h$ of 0.11~{\AA} for Cu(111) and 0.15~{\AA} for Ag(111), cf. Table~S~V\cite{SuppInf}). 
\renewcommand{\arraystretch}{1.2}
\begin{table}
 \centering
 \caption{Calculated workfunctions (``asymmetric slab``), values in parenthesis refer to calculations with dipole canceling slabs.}
\begin{tabular}{c|ccc}
\toprule
 & \num{5x5} & \num{6x6} &  \num{7x7}\\ \hline
Ag(111) & \ 3.97 eV \ & \ 4.06 eV (3.80 eV) \ & \ 4.12 eV \  \\ 
Cu(111) & 4.03 eV & 4.15 eV (3.81 eV) & 4.28 eV \\ \hline 
\end{tabular}
\label{tab:Workfunction}
\end{table}
\renewcommand{\arraystretch}{1}

On silver \mbox{2H-P} overlayers basically exhibit repulsion at high coverage and a fast decay towards a constant adsorption energy. This suggests that no energetic driving force for island formation exists, much in agreement with experimental findings. In the case of copper the initially repulsive contribution (stemming from the deformation energy) at high coverage is reduced and the \num{6x6} unit cell is found to be the most stable coverage at -3.16~eV. The stabilization of the \num{6x6} cell when compared to the low coverage limit is 0.01 eV, which can be considered small. In fact this corresponds to a temperature of 116 K. Therefore, not considering surface diffusion or anharmonicity or mode coupling effects~\cite{Mercurio2013}, below this energy a principal energetic driving force for self-assembly at an intermediate distance of 15.4~{\AA} exists. The thermodynamic picture is hence rather clear, in both cases, the driving forces for island assembly are non-existent or small. In fact, for both substrates the self-assembly into islands might be kinetically hindered by diffusion barriers between adsorption sites. Although we did not directly study the lateral diffusion process and the corresponding transition states, the significantly larger adsorption energy and deformation of \mbox{2H-P} on copper suggests that in this case such a barrier may be significant. This would lead to a kinetically hindered self-assembly process, where adsorbate molecules upon impingement on the surface are fixed at their initial adsorption site. Experiments for \mbox{2H-P} on copper suggest that this is in fact the case.~\cite{Diller2013}

\section{Conclusions and outlook}
\label{sec:conclusion}
Molecular self-assembly on surfaces is a major focus area in molecular nanotechnology and its detailed understanding in terms of individual binding contributions between adsorbate and substrate is an important goal. In this work we used dispersion-corrected density functional theory to provide such understanding for the prototypical case of free-base porphine adsorption on the coinage metal surfaces Ag(111) and Cu(111). We find that on both substrates the adsorbate-substrate interaction is dominated by attractive van der Waals forces. The adsorption is generally stronger on copper, leading to a higher deformation of the porphines, a higher binding energy and a more substantial modification of the frontier orbitals as evidenced by the MO-PDOS. The latter includes hybridization effects such as a strong broadening and splitting, as well as a down-shift of the LUMO and LUMO+1 (partially) below the Fermi level. This electron transfer plays a significant role only on the stronger interacting Cu(111) surface and is additionally confirmed by employing different charge partitioning schemes (Bader, Hirshfeld, Mulliken, integrated PDOS). Not unusual for organic adsorbates is the accompanying charge redistribution upon adsorption, manifesting as a pronounced pillow effect and charge accumulation around the molecules. By varying the surface unit-cell size and artificially switching on and off interactions between adsorbates and adsorbate and substrate, we were able to disentangle the various attractive and repulsive contributions to the adsorption energy for different coverages. Copper binds stronger than silver due to stronger van der Waals interaction and less DFT repulsion. Increasing coverage leads to an energetic penalty due to adsorbate deformation and to further stabilization due to substrate-mediated lateral dispersion interactions. The sum of the latter with the minimal amount of  intermolecular van der Waals attractions cannot outbalance the unfavorable deformation energy and DFT contribution at higher coverages. The resulting more unfavorable adsorption energy prevents the formation of molecular islands on both substrates. 

The reliability of the derived results depends of course on the quality of the computational method. While the adsorption geometry is typically well recovered with a standard semi-local functional like the employed PBE functional together with van der Waals corrections~\cite{Mercurio2013, Ruiz2012, Maurer2015} the combination of DFT with \emph{a posteriori} dispersion corrections can bias the description of the electronic structure~\cite{Ferri2015}. Our results agree in all points (adsorption on bridge sites, weak electron transfer on Ag, strong electron transfer on Cu, no island formation at high molecular coverages) with published interpretations from experiments,~\cite{Bischoff2013a,Diller2013}  and we do not have any indications for the occurrence of such systematic errors for our current case. Though, in general, great caution is advised when calculating density-derived observables. It is important to note that although the employed DFT+vdW$^\textrm{surf}$ approach only uses a pairwise-additive dispersion correction and thereby neglects many-body contributions to the dispersion energy~\cite{Tkatchenko2012, Maurer2015} the comparison with literature results obtained with van der Waals functionals~\cite{Dyer2011} and also with experiments~\cite{Bischoff2013a,Diller2013} strongly support our qualitative findings. Surprisingly, even small substrate-mediated lateral interactions due to changes in the electron density are captured, since the atomic polarizabilities $\alpha$ and C$_6$-coefficients of the vdW$^\textrm{surf}$ scheme are coupled to the density via Hirshfeld partitioning.

Overall, our results not only support existing experimental works and provide a first-principles understanding of interactions governing molecular self-assembly, but might also enable predications for the behavior of substituted porphyrins. By partitioning the adsorption energy (cf.\ Fig.\ \ref{fig:coverage}) for the basic porphine as starting point, we can try to access the anticipated outcome of functionalization by analyzing the influence on each single energy component. Adding, for example, phenyl side groups, introduces additional attractive van der Waals interactions at smaller molecule-molecule distances, which can be enough to facilitate the formation of islands on Ag(111).\cite{Rojas2012} Further works could include a detailed breakdown of the different energy components for molecular functionalization that targets electronic decoupling of the surface in order to reduce kinetic diffusion barriers\cite{Joshi2014} or to increase lateral interaction between adsorbates. Another interesting approach could  be to enhance substrate-mediated lateral interactions by modifying the substrate electronic structure, for example with electron-widhdrawing co-adsorbates.

\begin{acknowledgments}

Computer resources for this project have been provided by the Leibniz Supercomputing Center in Garching, grant pr85za. We thank Felix Bischoff and Willi Auw\"{a}rter for discussions. K.D. acknowledges support by the \'{E}cole polytechnique f\'{e}d\'{e}rale de Lausanne (EPFL).

\end{acknowledgments}


\begin{thebibliography}{58}%
\makeatletter
\providecommand \@ifxundefined [1]{%
 \@ifx{#1\undefined}
}%
\providecommand \@ifnum [1]{%
 \ifnum #1\expandafter \@firstoftwo
 \else \expandafter \@secondoftwo
 \fi
}%
\providecommand \@ifx [1]{%
 \ifx #1\expandafter \@firstoftwo
 \else \expandafter \@secondoftwo
 \fi
}%
\providecommand \natexlab [1]{#1}%
\providecommand \enquote  [1]{``#1''}%
\providecommand \bibnamefont  [1]{#1}%
\providecommand \bibfnamefont [1]{#1}%
\providecommand \citenamefont [1]{#1}%
\providecommand \href@noop [0]{\@secondoftwo}%
\providecommand \href [0]{\begingroup \@sanitize@url \@href}%
\providecommand \@href[1]{\@@startlink{#1}\@@href}%
\providecommand \@@href[1]{\endgroup#1\@@endlink}%
\providecommand \@sanitize@url [0]{\catcode `\\12\catcode `\$12\catcode
  `\&12\catcode `\#12\catcode `\^12\catcode `\_12\catcode `\%12\relax}%
\providecommand \@@startlink[1]{}%
\providecommand \@@endlink[0]{}%
\providecommand \url  [0]{\begingroup\@sanitize@url \@url }%
\providecommand \@url [1]{\endgroup\@href {#1}{\urlprefix }}%
\providecommand \urlprefix  [0]{URL }%
\providecommand \Eprint [0]{\href }%
\providecommand \doibase [0]{http://dx.doi.org/}%
\providecommand \selectlanguage [0]{\@gobble}%
\providecommand \bibinfo  [0]{\@secondoftwo}%
\providecommand \bibfield  [0]{\@secondoftwo}%
\providecommand \translation [1]{[#1]}%
\providecommand \BibitemOpen [0]{}%
\providecommand \bibitemStop [0]{}%
\providecommand \bibitemNoStop [0]{.\EOS\space}%
\providecommand \EOS [0]{\spacefactor3000\relax}%
\providecommand \BibitemShut  [1]{\csname bibitem#1\endcsname}%
\let\auto@bib@innerbib\@empty
\bibitem [{\citenamefont {Binnig}\ \emph {et~al.}(1982)\citenamefont {Binnig},
  \citenamefont {Rohrer}, \citenamefont {Gerber},\ and\ \citenamefont
  {Weibel}}]{Binnig1982}%
  \BibitemOpen
  \bibfield  {author} {\bibinfo {author} {\bibfnamefont {G.}~\bibnamefont
  {Binnig}}, \bibinfo {author} {\bibfnamefont {H.}~\bibnamefont {Rohrer}},
  \bibinfo {author} {\bibfnamefont {C.}~\bibnamefont {Gerber}}, \ and\ \bibinfo
  {author} {\bibfnamefont {E.}~\bibnamefont {Weibel}},\ }\href {\doibase
  10.1103/PhysRevLett.49.57} {\bibfield  {journal} {\bibinfo  {journal} {Phys.
  Rev. Lett.}\ }\textbf {\bibinfo {volume} {49}},\ \bibinfo {pages} {57}
  (\bibinfo {year} {1982})}\BibitemShut {NoStop}%
\bibitem [{\citenamefont {Li}\ \emph {et~al.}(2004)\citenamefont {Li},
  \citenamefont {Mathur}, \citenamefont {Gowda}, \citenamefont {Surthi},
  \citenamefont {Zhao}, \citenamefont {Yu}, \citenamefont {Lindsey},
  \citenamefont {Bocian},\ and\ \citenamefont {Misra}}]{Li2004}%
  \BibitemOpen
  \bibfield  {author} {\bibinfo {author} {\bibfnamefont {Q.}~\bibnamefont
  {Li}}, \bibinfo {author} {\bibfnamefont {G.}~\bibnamefont {Mathur}}, \bibinfo
  {author} {\bibfnamefont {S.}~\bibnamefont {Gowda}}, \bibinfo {author}
  {\bibfnamefont {S.}~\bibnamefont {Surthi}}, \bibinfo {author} {\bibfnamefont
  {Q.}~\bibnamefont {Zhao}}, \bibinfo {author} {\bibfnamefont {L.}~\bibnamefont
  {Yu}}, \bibinfo {author} {\bibfnamefont {J.~S.}\ \bibnamefont {Lindsey}},
  \bibinfo {author} {\bibfnamefont {D.~F.}\ \bibnamefont {Bocian}}, \ and\
  \bibinfo {author} {\bibfnamefont {V.}~\bibnamefont {Misra}},\ }\href
  {\doibase 10.1002/adma.200305680} {\bibfield  {journal} {\bibinfo  {journal}
  {Adv. Mater.}\ }\textbf {\bibinfo {volume} {16}},\ \bibinfo {pages} {133}
  (\bibinfo {year} {2004})}\BibitemShut {NoStop}%
\bibitem [{\citenamefont {Liu}\ \emph {et~al.}(2003)\citenamefont {Liu},
  \citenamefont {Yasseri}, \citenamefont {Lindsey},\ and\ \citenamefont
  {Bocian}}]{Liu2003}%
  \BibitemOpen
  \bibfield  {author} {\bibinfo {author} {\bibfnamefont {Z.}~\bibnamefont
  {Liu}}, \bibinfo {author} {\bibfnamefont {A.~A.}\ \bibnamefont {Yasseri}},
  \bibinfo {author} {\bibfnamefont {J.~S.}\ \bibnamefont {Lindsey}}, \ and\
  \bibinfo {author} {\bibfnamefont {D.~F.}\ \bibnamefont {Bocian}},\ }\href
  {\doibase 10.1126/science.1090677} {\bibfield  {journal} {\bibinfo  {journal}
  {Science}\ }\textbf {\bibinfo {volume} {302}},\ \bibinfo {pages} {1543}
  (\bibinfo {year} {2003})}\BibitemShut {NoStop}%
\bibitem [{\citenamefont {Campbell}\ \emph {et~al.}(2004)\citenamefont
  {Campbell}, \citenamefont {Burrell}, \citenamefont {Officer},\ and\
  \citenamefont {Jolley}}]{Campbell2004}%
  \BibitemOpen
  \bibfield  {author} {\bibinfo {author} {\bibfnamefont {W.~M.}\ \bibnamefont
  {Campbell}}, \bibinfo {author} {\bibfnamefont {A.~K.}\ \bibnamefont
  {Burrell}}, \bibinfo {author} {\bibfnamefont {D.~L.}\ \bibnamefont
  {Officer}}, \ and\ \bibinfo {author} {\bibfnamefont {K.~W.}\ \bibnamefont
  {Jolley}},\ }\href {\doibase 10.1016/j.ccr.2004.01.007} {\bibfield  {journal}
  {\bibinfo  {journal} {Coord. Chem. Rev.}\ }\textbf {\bibinfo {volume}
  {248}},\ \bibinfo {pages} {1363} (\bibinfo {year} {2004})}\BibitemShut
  {NoStop}%
\bibitem [{\citenamefont {Li}\ and\ \citenamefont {Diau}(2013)}]{Li2013a}%
  \BibitemOpen
  \bibfield  {author} {\bibinfo {author} {\bibfnamefont {L.-L.}\ \bibnamefont
  {Li}}\ and\ \bibinfo {author} {\bibfnamefont {E.~W.-G.}\ \bibnamefont
  {Diau}},\ }\href {\doibase 10.1039/c2cs35257e} {\bibfield  {journal}
  {\bibinfo  {journal} {Chem. Soc. Rev.}\ }\textbf {\bibinfo {volume} {42}},\
  \bibinfo {pages} {291} (\bibinfo {year} {2013})}\BibitemShut {NoStop}%
\bibitem [{\citenamefont {Rakow}\ and\ \citenamefont
  {Suslick}(2000)}]{Rakow2000}%
  \BibitemOpen
  \bibfield  {author} {\bibinfo {author} {\bibfnamefont {N.~A.}\ \bibnamefont
  {Rakow}}\ and\ \bibinfo {author} {\bibfnamefont {K.~S.}\ \bibnamefont
  {Suslick}},\ }\href {\doibase 10.1038/35021028} {\bibfield  {journal}
  {\bibinfo  {journal} {Nature}\ }\textbf {\bibinfo {volume} {406}},\ \bibinfo
  {pages} {710} (\bibinfo {year} {2000})}\BibitemShut {NoStop}%
\bibitem [{\citenamefont {Baldo}\ \emph {et~al.}(1998)\citenamefont {Baldo},
  \citenamefont {O'Brien}, \citenamefont {You}, \citenamefont {Shoustikov},
  \citenamefont {Sibley}, \citenamefont {Thompson},\ and\ \citenamefont
  {Forrest}}]{Baldo1998}%
  \BibitemOpen
  \bibfield  {author} {\bibinfo {author} {\bibfnamefont {M.~A.}\ \bibnamefont
  {Baldo}}, \bibinfo {author} {\bibfnamefont {D.~F.}\ \bibnamefont {O'Brien}},
  \bibinfo {author} {\bibfnamefont {Y.}~\bibnamefont {You}}, \bibinfo {author}
  {\bibfnamefont {A.}~\bibnamefont {Shoustikov}}, \bibinfo {author}
  {\bibfnamefont {S.}~\bibnamefont {Sibley}}, \bibinfo {author} {\bibfnamefont
  {M.~E.}\ \bibnamefont {Thompson}}, \ and\ \bibinfo {author} {\bibfnamefont
  {S.~R.}\ \bibnamefont {Forrest}},\ }\href {\doibase doi:10.1038/25954}
  {\bibfield  {journal} {\bibinfo  {journal} {Nature}\ }\textbf {\bibinfo
  {volume} {395}},\ \bibinfo {pages} {151} (\bibinfo {year}
  {1998})}\BibitemShut {NoStop}%
\bibitem [{\citenamefont {Su}\ \emph {et~al.}(2010)\citenamefont {Su},
  \citenamefont {Hatay}, \citenamefont {Troj\'{a}nek}, \citenamefont {Samec},
  \citenamefont {Khoury}, \citenamefont {Gros}, \citenamefont {Barbe},
  \citenamefont {Daina}, \citenamefont {Carrupt},\ and\ \citenamefont
  {Girault}}]{Su2010}%
  \BibitemOpen
  \bibfield  {author} {\bibinfo {author} {\bibfnamefont {B.}~\bibnamefont
  {Su}}, \bibinfo {author} {\bibfnamefont {I.}~\bibnamefont {Hatay}}, \bibinfo
  {author} {\bibfnamefont {A.}~\bibnamefont {Troj\'{a}nek}}, \bibinfo {author}
  {\bibfnamefont {Z.}~\bibnamefont {Samec}}, \bibinfo {author} {\bibfnamefont
  {T.}~\bibnamefont {Khoury}}, \bibinfo {author} {\bibfnamefont {C.~P.}\
  \bibnamefont {Gros}}, \bibinfo {author} {\bibfnamefont {J.-M.}\ \bibnamefont
  {Barbe}}, \bibinfo {author} {\bibfnamefont {A.}~\bibnamefont {Daina}},
  \bibinfo {author} {\bibfnamefont {P.-A.}\ \bibnamefont {Carrupt}}, \ and\
  \bibinfo {author} {\bibfnamefont {H.~H.}\ \bibnamefont {Girault}},\ }\href
  {\doibase 10.1021/ja908488s} {\bibfield  {journal} {\bibinfo  {journal} {J.
  Am. Chem. Soc.}\ }\textbf {\bibinfo {volume} {132}},\ \bibinfo {pages} {2655}
  (\bibinfo {year} {2010})}\BibitemShut {NoStop}%
\bibitem [{\citenamefont {Wintjes}\ \emph {et~al.}(2008)\citenamefont
  {Wintjes}, \citenamefont {Hornung}, \citenamefont {Lobo-Checa}, \citenamefont
  {Voigt}, \citenamefont {Samuely}, \citenamefont {Thilgen}, \citenamefont
  {St\"{o}hr}, \citenamefont {Diederich},\ and\ \citenamefont
  {Jung}}]{Wintjes2008}%
  \BibitemOpen
  \bibfield  {author} {\bibinfo {author} {\bibfnamefont {N.}~\bibnamefont
  {Wintjes}}, \bibinfo {author} {\bibfnamefont {J.}~\bibnamefont {Hornung}},
  \bibinfo {author} {\bibfnamefont {J.}~\bibnamefont {Lobo-Checa}}, \bibinfo
  {author} {\bibfnamefont {T.}~\bibnamefont {Voigt}}, \bibinfo {author}
  {\bibfnamefont {T.}~\bibnamefont {Samuely}}, \bibinfo {author} {\bibfnamefont
  {C.}~\bibnamefont {Thilgen}}, \bibinfo {author} {\bibfnamefont
  {M.}~\bibnamefont {St\"{o}hr}}, \bibinfo {author} {\bibfnamefont
  {F.}~\bibnamefont {Diederich}}, \ and\ \bibinfo {author} {\bibfnamefont
  {T.~A.}\ \bibnamefont {Jung}},\ }\href {\doibase 10.1002/chem.200800746}
  {\bibfield  {journal} {\bibinfo  {journal} {Chem. Eur. J.}\ }\textbf
  {\bibinfo {volume} {14}},\ \bibinfo {pages} {5794} (\bibinfo {year}
  {2008})}\BibitemShut {NoStop}%
\bibitem [{\citenamefont {Xiao}\ \emph {et~al.}(2012)\citenamefont {Xiao},
  \citenamefont {Ditze}, \citenamefont {Chen}, \citenamefont {Buchner},
  \citenamefont {Stark}, \citenamefont {Drost}, \citenamefont {Steinr\"{u}ck},
  \citenamefont {Gottfried},\ and\ \citenamefont {Marbach}}]{Xiao2012}%
  \BibitemOpen
  \bibfield  {author} {\bibinfo {author} {\bibfnamefont {J.}~\bibnamefont
  {Xiao}}, \bibinfo {author} {\bibfnamefont {S.}~\bibnamefont {Ditze}},
  \bibinfo {author} {\bibfnamefont {M.}~\bibnamefont {Chen}}, \bibinfo {author}
  {\bibfnamefont {F.}~\bibnamefont {Buchner}}, \bibinfo {author} {\bibfnamefont
  {M.}~\bibnamefont {Stark}}, \bibinfo {author} {\bibfnamefont
  {M.}~\bibnamefont {Drost}}, \bibinfo {author} {\bibfnamefont {H.-P.}\
  \bibnamefont {Steinr\"{u}ck}}, \bibinfo {author} {\bibfnamefont {J.~M.}\
  \bibnamefont {Gottfried}}, \ and\ \bibinfo {author} {\bibfnamefont
  {H.}~\bibnamefont {Marbach}},\ }\href {\doibase 10.1021/jp301757h} {\bibfield
   {journal} {\bibinfo  {journal} {J. Phys. Chem. C}\ }\textbf {\bibinfo
  {volume} {116}},\ \bibinfo {pages} {12275} (\bibinfo {year}
  {2012})}\BibitemShut {NoStop}%
\bibitem [{\citenamefont {Rojas}\ \emph {et~al.}(2010)\citenamefont {Rojas},
  \citenamefont {Chen}, \citenamefont {Bravo}, \citenamefont {Kim},
  \citenamefont {Kim}, \citenamefont {Xiao}, \citenamefont {Dowben},
  \citenamefont {Gao}, \citenamefont {Zeng}, \citenamefont {Choe},\ and\
  \citenamefont {Enders}}]{Rojas2010}%
  \BibitemOpen
  \bibfield  {author} {\bibinfo {author} {\bibfnamefont {G.}~\bibnamefont
  {Rojas}}, \bibinfo {author} {\bibfnamefont {X.}~\bibnamefont {Chen}},
  \bibinfo {author} {\bibfnamefont {C.}~\bibnamefont {Bravo}}, \bibinfo
  {author} {\bibfnamefont {J.-H.}\ \bibnamefont {Kim}}, \bibinfo {author}
  {\bibfnamefont {J.-S.}\ \bibnamefont {Kim}}, \bibinfo {author} {\bibfnamefont
  {J.}~\bibnamefont {Xiao}}, \bibinfo {author} {\bibfnamefont {P.~A.}\
  \bibnamefont {Dowben}}, \bibinfo {author} {\bibfnamefont {Y.}~\bibnamefont
  {Gao}}, \bibinfo {author} {\bibfnamefont {X.~C.}\ \bibnamefont {Zeng}},
  \bibinfo {author} {\bibfnamefont {W.}~\bibnamefont {Choe}}, \ and\ \bibinfo
  {author} {\bibfnamefont {A.}~\bibnamefont {Enders}},\ }\href {\doibase
  10.1021/jp1012957} {\bibfield  {journal} {\bibinfo  {journal} {J. Phys. Chem.
  C}\ }\textbf {\bibinfo {volume} {114}},\ \bibinfo {pages} {9408} (\bibinfo
  {year} {2010})}\BibitemShut {NoStop}%
\bibitem [{\citenamefont {Diller}\ \emph {et~al.}(2013)\citenamefont {Diller},
  \citenamefont {Klappenberger}, \citenamefont {Allegretti}, \citenamefont
  {Papageorgiou}, \citenamefont {Fischer}, \citenamefont {Wiengarten},
  \citenamefont {Joshi}, \citenamefont {Seufert}, \citenamefont {\'{E}cija},
  \citenamefont {Auw\"{a}rter},\ and\ \citenamefont {Barth}}]{Diller2013}%
  \BibitemOpen
  \bibfield  {author} {\bibinfo {author} {\bibfnamefont {K.}~\bibnamefont
  {Diller}}, \bibinfo {author} {\bibfnamefont {F.}~\bibnamefont
  {Klappenberger}}, \bibinfo {author} {\bibfnamefont {F.}~\bibnamefont
  {Allegretti}}, \bibinfo {author} {\bibfnamefont {A.~C.}\ \bibnamefont
  {Papageorgiou}}, \bibinfo {author} {\bibfnamefont {S.}~\bibnamefont
  {Fischer}}, \bibinfo {author} {\bibfnamefont {A.}~\bibnamefont {Wiengarten}},
  \bibinfo {author} {\bibfnamefont {S.}~\bibnamefont {Joshi}}, \bibinfo
  {author} {\bibfnamefont {K.}~\bibnamefont {Seufert}}, \bibinfo {author}
  {\bibfnamefont {D.}~\bibnamefont {\'{E}cija}}, \bibinfo {author}
  {\bibfnamefont {W.}~\bibnamefont {Auw\"{a}rter}}, \ and\ \bibinfo {author}
  {\bibfnamefont {J.~V.}\ \bibnamefont {Barth}},\ }\href {\doibase
  10.1063/1.4800771} {\bibfield  {journal} {\bibinfo  {journal} {J. Chem.
  Phys.}\ }\textbf {\bibinfo {volume} {138}},\ \bibinfo {pages} {154710}
  (\bibinfo {year} {2013})}\BibitemShut {NoStop}%
\bibitem [{\citenamefont {Webb}\ and\ \citenamefont
  {Fleischer}(1965)}]{Webb1965}%
  \BibitemOpen
  \bibfield  {author} {\bibinfo {author} {\bibfnamefont {L.~E.}\ \bibnamefont
  {Webb}}\ and\ \bibinfo {author} {\bibfnamefont {E.~B.}\ \bibnamefont
  {Fleischer}},\ }\href {\doibase 10.1063/1.1697283} {\bibfield  {journal}
  {\bibinfo  {journal} {J. Chem. Phys.}\ }\textbf {\bibinfo {volume} {43}},\
  \bibinfo {pages} {3100} (\bibinfo {year} {1965})}\BibitemShut {NoStop}%
\bibitem [{\citenamefont {Ort\'{\i}}\ and\ \citenamefont
  {Br\'{e}das}(1989)}]{Orti1989}%
  \BibitemOpen
  \bibfield  {author} {\bibinfo {author} {\bibfnamefont {E.}~\bibnamefont
  {Ort\'{\i}}}\ and\ \bibinfo {author} {\bibfnamefont {J.~L.}\ \bibnamefont
  {Br\'{e}das}},\ }\href {\doibase 10.1016/0009-2614(89)85023-7} {\bibfield
  {journal} {\bibinfo  {journal} {Chem. Phys. Lett.}\ }\textbf {\bibinfo
  {volume} {164}},\ \bibinfo {pages} {247} (\bibinfo {year}
  {1989})}\BibitemShut {NoStop}%
\bibitem [{\citenamefont {Blase}, \citenamefont {Attaccalite},\ and\
  \citenamefont {Olevano}(2011)}]{Blase2011a}%
  \BibitemOpen
  \bibfield  {author} {\bibinfo {author} {\bibfnamefont {X.}~\bibnamefont
  {Blase}}, \bibinfo {author} {\bibfnamefont {C.}~\bibnamefont {Attaccalite}},
  \ and\ \bibinfo {author} {\bibfnamefont {V.}~\bibnamefont {Olevano}},\ }\href
  {\doibase 10.1103/PhysRevB.83.115103} {\bibfield  {journal} {\bibinfo
  {journal} {Phys. Rev. B}\ }\textbf {\bibinfo {volume} {83}},\ \bibinfo
  {pages} {115103} (\bibinfo {year} {2011})}\BibitemShut {NoStop}%
\bibitem [{\citenamefont {Dabo}\ \emph {et~al.}(2013)\citenamefont {Dabo},
  \citenamefont {Ferretti}, \citenamefont {Park}, \citenamefont {Poilvert},
  \citenamefont {Li}, \citenamefont {Cococcioni},\ and\ \citenamefont
  {Marzari}}]{Dabo2013}%
  \BibitemOpen
  \bibfield  {author} {\bibinfo {author} {\bibfnamefont {I.}~\bibnamefont
  {Dabo}}, \bibinfo {author} {\bibfnamefont {A.}~\bibnamefont {Ferretti}},
  \bibinfo {author} {\bibfnamefont {C.-H.}\ \bibnamefont {Park}}, \bibinfo
  {author} {\bibfnamefont {N.}~\bibnamefont {Poilvert}}, \bibinfo {author}
  {\bibfnamefont {Y.}~\bibnamefont {Li}}, \bibinfo {author} {\bibfnamefont
  {M.}~\bibnamefont {Cococcioni}}, \ and\ \bibinfo {author} {\bibfnamefont
  {N.}~\bibnamefont {Marzari}},\ }\href {\doibase 10.1039/c2cp43491a}
  {\bibfield  {journal} {\bibinfo  {journal} {Phys. Chem. Chem. Phys.}\
  }\textbf {\bibinfo {volume} {15}},\ \bibinfo {pages} {685} (\bibinfo {year}
  {2013})}\BibitemShut {NoStop}%
\bibitem [{\citenamefont {Zhang}\ \emph {et~al.}(2005)\citenamefont {Zhang},
  \citenamefont {Li}, \citenamefont {Wu}, \citenamefont {Zhu},\ and\
  \citenamefont {Zheng}}]{Zhang2005}%
  \BibitemOpen
  \bibfield  {author} {\bibinfo {author} {\bibfnamefont {Y.-H.}\ \bibnamefont
  {Zhang}}, \bibinfo {author} {\bibfnamefont {Z.-Y.}\ \bibnamefont {Li}},
  \bibinfo {author} {\bibfnamefont {Y.}~\bibnamefont {Wu}}, \bibinfo {author}
  {\bibfnamefont {Y.-Z.}\ \bibnamefont {Zhu}}, \ and\ \bibinfo {author}
  {\bibfnamefont {J.-Y.}\ \bibnamefont {Zheng}},\ }\href {\doibase
  10.1016/j.saa.2004.12.009} {\bibfield  {journal} {\bibinfo  {journal}
  {Spectrochim. Acta, Part A}\ }\textbf {\bibinfo {volume} {62}},\ \bibinfo
  {pages} {83} (\bibinfo {year} {2005})}\BibitemShut {NoStop}%
\bibitem [{\citenamefont {Haq}\ \emph {et~al.}(2011)\citenamefont {Haq},
  \citenamefont {Hanke}, \citenamefont {Dyer}, \citenamefont {Persson},
  \citenamefont {Iavicoli}, \citenamefont {Amabilino},\ and\ \citenamefont
  {Raval}}]{Haq2011}%
  \BibitemOpen
  \bibfield  {author} {\bibinfo {author} {\bibfnamefont {S.}~\bibnamefont
  {Haq}}, \bibinfo {author} {\bibfnamefont {F.}~\bibnamefont {Hanke}}, \bibinfo
  {author} {\bibfnamefont {M.~S.}\ \bibnamefont {Dyer}}, \bibinfo {author}
  {\bibfnamefont {M.}~\bibnamefont {Persson}}, \bibinfo {author} {\bibfnamefont
  {P.}~\bibnamefont {Iavicoli}}, \bibinfo {author} {\bibfnamefont {D.~B.}\
  \bibnamefont {Amabilino}}, \ and\ \bibinfo {author} {\bibfnamefont
  {R.}~\bibnamefont {Raval}},\ }\href {\doibase 10.1021/ja201389u} {\bibfield
  {journal} {\bibinfo  {journal} {J. Am. Chem. Soc.}\ }\textbf {\bibinfo
  {volume} {133}},\ \bibinfo {pages} {12031} (\bibinfo {year}
  {2011})}\BibitemShut {NoStop}%
\bibitem [{\citenamefont {Iacovita}\ \emph {et~al.}(2012)\citenamefont
  {Iacovita}, \citenamefont {Fesser}, \citenamefont {Vijayaraghavan},
  \citenamefont {Enache}, \citenamefont {St\"{o}hr}, \citenamefont
  {Diederich},\ and\ \citenamefont {Jung}}]{Iacovita2012}%
  \BibitemOpen
  \bibfield  {author} {\bibinfo {author} {\bibfnamefont {C.}~\bibnamefont
  {Iacovita}}, \bibinfo {author} {\bibfnamefont {P.}~\bibnamefont {Fesser}},
  \bibinfo {author} {\bibfnamefont {S.}~\bibnamefont {Vijayaraghavan}},
  \bibinfo {author} {\bibfnamefont {M.}~\bibnamefont {Enache}}, \bibinfo
  {author} {\bibfnamefont {M.}~\bibnamefont {St\"{o}hr}}, \bibinfo {author}
  {\bibfnamefont {F.}~\bibnamefont {Diederich}}, \ and\ \bibinfo {author}
  {\bibfnamefont {T.~A.}\ \bibnamefont {Jung}},\ }\href {\doibase
  10.1002/chem.201201037} {\bibfield  {journal} {\bibinfo  {journal} {Chem.
  Eur. J.}\ }\textbf {\bibinfo {volume} {18}},\ \bibinfo {pages} {14610}
  (\bibinfo {year} {2012})}\BibitemShut {NoStop}%
\bibitem [{\citenamefont {Bischoff}\ \emph {et~al.}(2013)\citenamefont
  {Bischoff}, \citenamefont {Seufert}, \citenamefont {Auw\"{a}rter},
  \citenamefont {Joshi}, \citenamefont {Vijayaraghavan}, \citenamefont
  {\'{E}cija}, \citenamefont {Diller}, \citenamefont {Papageorgiou},
  \citenamefont {Fischer}, \citenamefont {Allegretti}, \citenamefont {Duncan},
  \citenamefont {Klappenberger}, \citenamefont {Blobner}, \citenamefont {Han},\
  and\ \citenamefont {Barth}}]{Bischoff2013a}%
  \BibitemOpen
  \bibfield  {author} {\bibinfo {author} {\bibfnamefont {F.}~\bibnamefont
  {Bischoff}}, \bibinfo {author} {\bibfnamefont {K.}~\bibnamefont {Seufert}},
  \bibinfo {author} {\bibfnamefont {W.}~\bibnamefont {Auw\"{a}rter}}, \bibinfo
  {author} {\bibfnamefont {S.}~\bibnamefont {Joshi}}, \bibinfo {author}
  {\bibfnamefont {S.}~\bibnamefont {Vijayaraghavan}}, \bibinfo {author}
  {\bibfnamefont {D.}~\bibnamefont {\'{E}cija}}, \bibinfo {author}
  {\bibfnamefont {K.}~\bibnamefont {Diller}}, \bibinfo {author} {\bibfnamefont
  {A.~C.}\ \bibnamefont {Papageorgiou}}, \bibinfo {author} {\bibfnamefont
  {S.}~\bibnamefont {Fischer}}, \bibinfo {author} {\bibfnamefont
  {F.}~\bibnamefont {Allegretti}}, \bibinfo {author} {\bibfnamefont {D.~A.}\
  \bibnamefont {Duncan}}, \bibinfo {author} {\bibfnamefont {F.}~\bibnamefont
  {Klappenberger}}, \bibinfo {author} {\bibfnamefont {F.}~\bibnamefont
  {Blobner}}, \bibinfo {author} {\bibfnamefont {R.}~\bibnamefont {Han}}, \ and\
  \bibinfo {author} {\bibfnamefont {J.~V.}\ \bibnamefont {Barth}},\ }\href
  {\doibase 10.1021/nn305487c} {\bibfield  {journal} {\bibinfo  {journal} {ACS
  Nano}\ }\textbf {\bibinfo {volume} {7}},\ \bibinfo {pages} {3139} (\bibinfo
  {year} {2013})}\BibitemShut {NoStop}%
\bibitem [{\citenamefont {Segall}\ \emph {et~al.}(2002)\citenamefont {Segall},
  \citenamefont {Lindan}, \citenamefont {Probert}, \citenamefont {Pickard},
  \citenamefont {Hasnip}, \citenamefont {Clark},\ and\ \citenamefont
  {Payne}}]{Segall2002}%
  \BibitemOpen
  \bibfield  {author} {\bibinfo {author} {\bibfnamefont {M.~D.}\ \bibnamefont
  {Segall}}, \bibinfo {author} {\bibfnamefont {P.~J.~D.}\ \bibnamefont
  {Lindan}}, \bibinfo {author} {\bibfnamefont {M.~J.}\ \bibnamefont {Probert}},
  \bibinfo {author} {\bibfnamefont {C.~J.}\ \bibnamefont {Pickard}}, \bibinfo
  {author} {\bibfnamefont {P.~J.}\ \bibnamefont {Hasnip}}, \bibinfo {author}
  {\bibfnamefont {S.~J.}\ \bibnamefont {Clark}}, \ and\ \bibinfo {author}
  {\bibfnamefont {M.~C.}\ \bibnamefont {Payne}},\ }\href {\doibase
  10.1088/0953-8984/14/11/301} {\bibfield  {journal} {\bibinfo  {journal} {J.
  Phys. Condens. Matter}\ }\textbf {\bibinfo {volume} {14}},\ \bibinfo {pages}
  {2717} (\bibinfo {year} {2002})}\BibitemShut {NoStop}%
\bibitem [{\citenamefont {Payne}\ \emph {et~al.}(1992)\citenamefont {Payne},
  \citenamefont {Teter}, \citenamefont {Allan}, \citenamefont {Arias},\ and\
  \citenamefont {Joannopoulos}}]{Payne1992}%
  \BibitemOpen
  \bibfield  {author} {\bibinfo {author} {\bibfnamefont {M.~C.}\ \bibnamefont
  {Payne}}, \bibinfo {author} {\bibfnamefont {M.~P.}\ \bibnamefont {Teter}},
  \bibinfo {author} {\bibfnamefont {D.~C.}\ \bibnamefont {Allan}}, \bibinfo
  {author} {\bibfnamefont {T.~A.}\ \bibnamefont {Arias}}, \ and\ \bibinfo
  {author} {\bibfnamefont {J.~D.}\ \bibnamefont {Joannopoulos}},\ }\href
  {\doibase 10.1103/RevModPhys.64.1045} {\bibfield  {journal} {\bibinfo
  {journal} {Rev. Mod. Phys.}\ }\textbf {\bibinfo {volume} {64}},\ \bibinfo
  {pages} {1045} (\bibinfo {year} {1992})}\BibitemShut {NoStop}%
\bibitem [{\citenamefont {Clark}\ \emph {et~al.}(2005)\citenamefont {Clark},
  \citenamefont {Segall}, \citenamefont {Pickard}, \citenamefont {Hasnip},
  \citenamefont {Probert}, \citenamefont {Refson},\ and\ \citenamefont
  {Payne}}]{Clark2005}%
  \BibitemOpen
  \bibfield  {author} {\bibinfo {author} {\bibfnamefont {S.~J.}\ \bibnamefont
  {Clark}}, \bibinfo {author} {\bibfnamefont {M.~D.}\ \bibnamefont {Segall}},
  \bibinfo {author} {\bibfnamefont {C.~J.}\ \bibnamefont {Pickard}}, \bibinfo
  {author} {\bibfnamefont {P.~J.}\ \bibnamefont {Hasnip}}, \bibinfo {author}
  {\bibfnamefont {M.~I.~J.}\ \bibnamefont {Probert}}, \bibinfo {author}
  {\bibfnamefont {K.}~\bibnamefont {Refson}}, \ and\ \bibinfo {author}
  {\bibfnamefont {M.~C.}\ \bibnamefont {Payne}},\ }\href {\doibase
  10.1524/zkri.220.5.567.65075} {\bibfield  {journal} {\bibinfo  {journal} {Z.
  Kristallogr.}\ }\textbf {\bibinfo {volume} {220}},\ \bibinfo {pages} {567}
  (\bibinfo {year} {2005})}\BibitemShut {NoStop}%
\bibitem [{\citenamefont {Vanderbilt}(1990)}]{Vanderbilt1990}%
  \BibitemOpen
  \bibfield  {author} {\bibinfo {author} {\bibfnamefont {D.}~\bibnamefont
  {Vanderbilt}},\ }\href {\doibase 10.1103/PhysRevB.41.7892} {\bibfield
  {journal} {\bibinfo  {journal} {Phys. Rev. B}\ }\textbf {\bibinfo {volume}
  {41}},\ \bibinfo {pages} {7892} (\bibinfo {year} {1990})}\BibitemShut
  {NoStop}%
\bibitem [{\citenamefont {Perdew}, \citenamefont {Burke},\ and\ \citenamefont
  {Ernzerhof}(1996)}]{Perdew1996}%
  \BibitemOpen
  \bibfield  {author} {\bibinfo {author} {\bibfnamefont {J.~P.}\ \bibnamefont
  {Perdew}}, \bibinfo {author} {\bibfnamefont {K.}~\bibnamefont {Burke}}, \
  and\ \bibinfo {author} {\bibfnamefont {M.}~\bibnamefont {Ernzerhof}},\ }\href
  {\doibase 10.1103/PhysRevLett.77.3865} {\bibfield  {journal} {\bibinfo
  {journal} {Phys. Rev. Lett.}\ }\textbf {\bibinfo {volume} {77}},\ \bibinfo
  {pages} {3865} (\bibinfo {year} {1996})}\BibitemShut {NoStop}%
\bibitem [{\citenamefont {Ruiz}\ \emph {et~al.}(2012)\citenamefont {Ruiz},
  \citenamefont {Liu}, \citenamefont {Zojer}, \citenamefont {Scheffler},\ and\
  \citenamefont {Tkatchenko}}]{Ruiz2012}%
  \BibitemOpen
  \bibfield  {author} {\bibinfo {author} {\bibfnamefont {V.~G.}\ \bibnamefont
  {Ruiz}}, \bibinfo {author} {\bibfnamefont {W.}~\bibnamefont {Liu}}, \bibinfo
  {author} {\bibfnamefont {E.}~\bibnamefont {Zojer}}, \bibinfo {author}
  {\bibfnamefont {M.}~\bibnamefont {Scheffler}}, \ and\ \bibinfo {author}
  {\bibfnamefont {A.}~\bibnamefont {Tkatchenko}},\ }\href {\doibase
  10.1103/PhysRevLett.108.146103} {\bibfield  {journal} {\bibinfo  {journal}
  {Phys. Rev. Lett.}\ }\textbf {\bibinfo {volume} {108}},\ \bibinfo {pages}
  {146103} (\bibinfo {year} {2012})}\BibitemShut {NoStop}%
\bibitem [{\citenamefont {Tkatchenko}\ and\ \citenamefont
  {Scheffler}(2009)}]{Tkatchenko2009}%
  \BibitemOpen
  \bibfield  {author} {\bibinfo {author} {\bibfnamefont {A.}~\bibnamefont
  {Tkatchenko}}\ and\ \bibinfo {author} {\bibfnamefont {M.}~\bibnamefont
  {Scheffler}},\ }\href {\doibase 10.1103/PhysRevLett.102.073005} {\bibfield
  {journal} {\bibinfo  {journal} {Phys. Rev. Lett.}\ }\textbf {\bibinfo
  {volume} {102}},\ \bibinfo {pages} {073005} (\bibinfo {year}
  {2009})}\BibitemShut {NoStop}%
\bibitem [{\citenamefont {Hofmann}\ \emph {et~al.}(2013)\citenamefont
  {Hofmann}, \citenamefont {Atalla}, \citenamefont {Moll}, \citenamefont
  {Rinke},\ and\ \citenamefont {Scheffler}}]{Hofmann2013}%
  \BibitemOpen
  \bibfield  {author} {\bibinfo {author} {\bibfnamefont {O.~T.}\ \bibnamefont
  {Hofmann}}, \bibinfo {author} {\bibfnamefont {V.}~\bibnamefont {Atalla}},
  \bibinfo {author} {\bibfnamefont {N.}~\bibnamefont {Moll}}, \bibinfo {author}
  {\bibfnamefont {P.}~\bibnamefont {Rinke}}, \ and\ \bibinfo {author}
  {\bibfnamefont {M.}~\bibnamefont {Scheffler}},\ }\href {\doibase
  10.1088/1367-2630/15/12/123028} {\bibfield  {journal} {\bibinfo  {journal}
  {New J. Phys.}\ }\textbf {\bibinfo {volume} {15}},\ \bibinfo {pages} {123028}
  (\bibinfo {year} {2013})}\BibitemShut {NoStop}%
\bibitem [{\citenamefont {Bahn}\ and\ \citenamefont
  {Jacobsen}(2002)}]{Bahn2002}%
  \BibitemOpen
  \bibfield  {author} {\bibinfo {author} {\bibfnamefont {S.}~\bibnamefont
  {Bahn}}\ and\ \bibinfo {author} {\bibfnamefont {K.}~\bibnamefont
  {Jacobsen}},\ }\href {\doibase 10.1109/5992.998641} {\bibfield  {journal}
  {\bibinfo  {journal} {Comput. Sci. Eng.}\ }\textbf {\bibinfo {volume} {4}},\
  \bibinfo {pages} {56} (\bibinfo {year} {2002})}\BibitemShut {NoStop}%
\bibitem [{\citenamefont {Hanke}\ \emph {et~al.}(2011)\citenamefont {Hanke},
  \citenamefont {Haq}, \citenamefont {Raval},\ and\ \citenamefont
  {Persson}}]{Hanke2011}%
  \BibitemOpen
  \bibfield  {author} {\bibinfo {author} {\bibfnamefont {F.}~\bibnamefont
  {Hanke}}, \bibinfo {author} {\bibfnamefont {S.}~\bibnamefont {Haq}}, \bibinfo
  {author} {\bibfnamefont {R.}~\bibnamefont {Raval}}, \ and\ \bibinfo {author}
  {\bibfnamefont {M.}~\bibnamefont {Persson}},\ }\href {\doibase
  10.1021/nn203337v} {\bibfield  {journal} {\bibinfo  {journal} {ACS Nano}\
  }\textbf {\bibinfo {volume} {5}},\ \bibinfo {pages} {9093} (\bibinfo {year}
  {2011})}\BibitemShut {NoStop}%
\bibitem [{\citenamefont {Dyer}\ \emph {et~al.}(2011)\citenamefont {Dyer},
  \citenamefont {Robin}, \citenamefont {Haq}, \citenamefont {Raval},
  \citenamefont {Persson},\ and\ \citenamefont {Klimes}}]{Dyer2011}%
  \BibitemOpen
  \bibfield  {author} {\bibinfo {author} {\bibfnamefont {M.~S.}\ \bibnamefont
  {Dyer}}, \bibinfo {author} {\bibfnamefont {A.}~\bibnamefont {Robin}},
  \bibinfo {author} {\bibfnamefont {S.}~\bibnamefont {Haq}}, \bibinfo {author}
  {\bibfnamefont {R.}~\bibnamefont {Raval}}, \bibinfo {author} {\bibfnamefont
  {M.}~\bibnamefont {Persson}}, \ and\ \bibinfo {author} {\bibfnamefont
  {J.}~\bibnamefont {Klimes}},\ }\href {\doibase 10.1021/nn102610k} {\bibfield
  {journal} {\bibinfo  {journal} {ACS Nano}\ }\textbf {\bibinfo {volume} {5}},\
  \bibinfo {pages} {1831} (\bibinfo {year} {2011})}\BibitemShut {NoStop}%
\bibitem [{\citenamefont {Monkhorst}\ and\ \citenamefont
  {Pack}(1976)}]{Monkhorst1976}%
  \BibitemOpen
  \bibfield  {author} {\bibinfo {author} {\bibfnamefont {H.~J.}\ \bibnamefont
  {Monkhorst}}\ and\ \bibinfo {author} {\bibfnamefont {J.~D.}\ \bibnamefont
  {Pack}},\ }\href {\doibase 10.1103/PhysRevB.13.5188} {\bibfield  {journal}
  {\bibinfo  {journal} {Phys. Rev. B}\ }\textbf {\bibinfo {volume} {13}},\
  \bibinfo {pages} {5188} (\bibinfo {year} {1976})}\BibitemShut {NoStop}%
\bibitem [{\citenamefont {Andzelm}, \citenamefont {King-Smith},\ and\
  \citenamefont {Fitzgerald}(2001)}]{Andzelm2001}%
  \BibitemOpen
  \bibfield  {author} {\bibinfo {author} {\bibfnamefont {J.}~\bibnamefont
  {Andzelm}}, \bibinfo {author} {\bibfnamefont {R.}~\bibnamefont {King-Smith}},
  \ and\ \bibinfo {author} {\bibfnamefont {G.}~\bibnamefont {Fitzgerald}},\
  }\href {\doibase 10.1016/S0009-2614(01)00030-6} {\bibfield  {journal}
  {\bibinfo  {journal} {Chem. Phys. Lett.}\ }\textbf {\bibinfo {volume}
  {335}},\ \bibinfo {pages} {321} (\bibinfo {year} {2001})}\BibitemShut
  {NoStop}%
\bibitem [{\citenamefont {Bader}(1991)}]{Bader1991}%
  \BibitemOpen
  \bibfield  {author} {\bibinfo {author} {\bibfnamefont {R.~F.~W.}\
  \bibnamefont {Bader}},\ }\href {\doibase 10.1021/cr00005a013} {\bibfield
  {journal} {\bibinfo  {journal} {Chem. Rev.}\ }\textbf {\bibinfo {volume}
  {91}},\ \bibinfo {pages} {893} (\bibinfo {year} {1991})}\BibitemShut
  {NoStop}%
\bibitem [{\citenamefont {Bader}(1994)}]{Bader1994}%
  \BibitemOpen
  \bibfield  {author} {\bibinfo {author} {\bibfnamefont {R.~F.~W.}\
  \bibnamefont {Bader}},\ }\href@noop {} {\emph {\bibinfo {title} {{Atoms in
  Molecules: A Quantum Theory}}}},\ \bibinfo {edition} {2nd}\ ed.\ (\bibinfo
  {address} {Oxford},\ \bibinfo {year} {1994})\BibitemShut {NoStop}%
\bibitem [{\citenamefont {Mulliken}(1955{\natexlab{a}})}]{Mulliken1955c}%
  \BibitemOpen
  \bibfield  {author} {\bibinfo {author} {\bibfnamefont {R.~S.}\ \bibnamefont
  {Mulliken}},\ }\href {\doibase 10.1063/1.1740588} {\bibfield  {journal}
  {\bibinfo  {journal} {J. Chem. Phys.}\ }\textbf {\bibinfo {volume} {23}},\
  \bibinfo {pages} {1833} (\bibinfo {year} {1955}{\natexlab{a}})}\BibitemShut
  {NoStop}%
\bibitem [{\citenamefont {Mulliken}(1955{\natexlab{b}})}]{Mulliken1955b}%
  \BibitemOpen
  \bibfield  {author} {\bibinfo {author} {\bibfnamefont {R.~S.}\ \bibnamefont
  {Mulliken}},\ }\href {\doibase 10.1063/1.1740589} {\bibfield  {journal}
  {\bibinfo  {journal} {J. Chem. Phys.}\ }\textbf {\bibinfo {volume} {23}},\
  \bibinfo {pages} {1841} (\bibinfo {year} {1955}{\natexlab{b}})}\BibitemShut
  {NoStop}%
\bibitem [{\citenamefont {Mulliken}(1955{\natexlab{c}})}]{Mulliken1955a}%
  \BibitemOpen
  \bibfield  {author} {\bibinfo {author} {\bibfnamefont {R.~S.}\ \bibnamefont
  {Mulliken}},\ }\href {\doibase 10.1063/1.1741876} {\bibfield  {journal}
  {\bibinfo  {journal} {J. Chem. Phys.}\ }\textbf {\bibinfo {volume} {23}},\
  \bibinfo {pages} {2338} (\bibinfo {year} {1955}{\natexlab{c}})}\BibitemShut
  {NoStop}%
\bibitem [{\citenamefont {Mulliken}(1955{\natexlab{d}})}]{Mulliken1955}%
  \BibitemOpen
  \bibfield  {author} {\bibinfo {author} {\bibfnamefont {R.~S.}\ \bibnamefont
  {Mulliken}},\ }\href {\doibase 10.1063/1.1741877} {\bibfield  {journal}
  {\bibinfo  {journal} {J. Chem. Phys.}\ }\textbf {\bibinfo {volume} {23}},\
  \bibinfo {pages} {2343} (\bibinfo {year} {1955}{\natexlab{d}})}\BibitemShut
  {NoStop}%
\bibitem [{\citenamefont {Sanchez-Portal}, \citenamefont {Artacho},\ and\
  \citenamefont {Soler}(1995)}]{Sanchez-Portal1995}%
  \BibitemOpen
  \bibfield  {author} {\bibinfo {author} {\bibfnamefont {D.}~\bibnamefont
  {Sanchez-Portal}}, \bibinfo {author} {\bibfnamefont {E.}~\bibnamefont
  {Artacho}}, \ and\ \bibinfo {author} {\bibfnamefont {J.~M.}\ \bibnamefont
  {Soler}},\ }\href {\doibase 10.1016/0038-1098(95)00341-X} {\bibfield
  {journal} {\bibinfo  {journal} {Solid State Commun.}\ }\textbf {\bibinfo
  {volume} {95}},\ \bibinfo {pages} {685} (\bibinfo {year} {1995})}\BibitemShut
  {NoStop}%
\bibitem [{\citenamefont {Hirshfeld}(1977)}]{Hirshfeld1977}%
  \BibitemOpen
  \bibfield  {author} {\bibinfo {author} {\bibfnamefont {F.~L.}\ \bibnamefont
  {Hirshfeld}},\ }\href {\doibase 10.1007/BF00549096} {\bibfield  {journal}
  {\bibinfo  {journal} {Theor. Chim. Acta}\ }\textbf {\bibinfo {volume} {44}},\
  \bibinfo {pages} {129} (\bibinfo {year} {1977})}\BibitemShut {NoStop}%
\bibitem [{\citenamefont {Henkelman}, \citenamefont {Arnaldsson},\ and\
  \citenamefont {J\'{o}nsson}(2006)}]{Henkelman2006}%
  \BibitemOpen
  \bibfield  {author} {\bibinfo {author} {\bibfnamefont {G.}~\bibnamefont
  {Henkelman}}, \bibinfo {author} {\bibfnamefont {A.}~\bibnamefont
  {Arnaldsson}}, \ and\ \bibinfo {author} {\bibfnamefont {H.}~\bibnamefont
  {J\'{o}nsson}},\ }\href {\doibase 10.1016/j.commatsci.2005.04.010} {\bibfield
   {journal} {\bibinfo  {journal} {Comput. Mater. Sci.}\ }\textbf {\bibinfo
  {volume} {36}},\ \bibinfo {pages} {354} (\bibinfo {year} {2006})}\BibitemShut
  {NoStop}%
\bibitem [{\citenamefont {Sanville}\ \emph {et~al.}(2007)\citenamefont
  {Sanville}, \citenamefont {Kenny}, \citenamefont {Smith},\ and\ \citenamefont
  {Henkelman}}]{Sanville2007}%
  \BibitemOpen
  \bibfield  {author} {\bibinfo {author} {\bibfnamefont {E.}~\bibnamefont
  {Sanville}}, \bibinfo {author} {\bibfnamefont {S.~D.}\ \bibnamefont {Kenny}},
  \bibinfo {author} {\bibfnamefont {R.}~\bibnamefont {Smith}}, \ and\ \bibinfo
  {author} {\bibfnamefont {G.}~\bibnamefont {Henkelman}},\ }\href {\doibase
  10.1002/jcc.20575} {\bibfield  {journal} {\bibinfo  {journal} {J. Comput.
  Chem.}\ }\textbf {\bibinfo {volume} {28}},\ \bibinfo {pages} {899} (\bibinfo
  {year} {2007})}\BibitemShut {NoStop}%
\bibitem [{\citenamefont {Tang}, \citenamefont {Sanville},\ and\ \citenamefont
  {Henkelman}(2009)}]{Tang2009}%
  \BibitemOpen
  \bibfield  {author} {\bibinfo {author} {\bibfnamefont {W.}~\bibnamefont
  {Tang}}, \bibinfo {author} {\bibfnamefont {E.}~\bibnamefont {Sanville}}, \
  and\ \bibinfo {author} {\bibfnamefont {G.}~\bibnamefont {Henkelman}},\ }\href
  {\doibase 10.1088/0953-8984/21/8/084204} {\bibfield  {journal} {\bibinfo
  {journal} {J. Phys. Condens. Matter}\ }\textbf {\bibinfo {volume} {21}},\
  \bibinfo {pages} {084204} (\bibinfo {year} {2009})}\BibitemShut {NoStop}%
\bibitem [{\citenamefont {Diller}\ \emph {et~al.}(2014)\citenamefont {Diller},
  \citenamefont {Klappenberger}, \citenamefont {Allegretti}, \citenamefont
  {Papageorgiou}, \citenamefont {Fischer}, \citenamefont {Duncan},
  \citenamefont {Maurer}, \citenamefont {Lloyd}, \citenamefont {Oh},
  \citenamefont {Reuter},\ and\ \citenamefont {Barth}}]{Diller2014}%
  \BibitemOpen
  \bibfield  {author} {\bibinfo {author} {\bibfnamefont {K.}~\bibnamefont
  {Diller}}, \bibinfo {author} {\bibfnamefont {F.}~\bibnamefont
  {Klappenberger}}, \bibinfo {author} {\bibfnamefont {F.}~\bibnamefont
  {Allegretti}}, \bibinfo {author} {\bibfnamefont {A.~C.}\ \bibnamefont
  {Papageorgiou}}, \bibinfo {author} {\bibfnamefont {S.}~\bibnamefont
  {Fischer}}, \bibinfo {author} {\bibfnamefont {D.~A.}\ \bibnamefont {Duncan}},
  \bibinfo {author} {\bibfnamefont {R.~J.}\ \bibnamefont {Maurer}}, \bibinfo
  {author} {\bibfnamefont {J.~A.}\ \bibnamefont {Lloyd}}, \bibinfo {author}
  {\bibfnamefont {S.~C.}\ \bibnamefont {Oh}}, \bibinfo {author} {\bibfnamefont
  {K.}~\bibnamefont {Reuter}}, \ and\ \bibinfo {author} {\bibfnamefont {J.~V.}\
  \bibnamefont {Barth}},\ }\href {\doibase 10.1063/1.4896605} {\bibfield
  {journal} {\bibinfo  {journal} {J. Chem. Phys.}\ }\textbf {\bibinfo {volume}
  {141}},\ \bibinfo {pages} {144703} (\bibinfo {year} {2014})}\BibitemShut
  {NoStop}%
\bibitem [{\citenamefont {Rojas}\ \emph {et~al.}(2012)\citenamefont {Rojas},
  \citenamefont {Simpson}, \citenamefont {Chen}, \citenamefont {Kunkel},
  \citenamefont {Nitz}, \citenamefont {Xiao}, \citenamefont {Dowben},
  \citenamefont {Zurek},\ and\ \citenamefont {Enders}}]{Rojas2012}%
  \BibitemOpen
  \bibfield  {author} {\bibinfo {author} {\bibfnamefont {G.}~\bibnamefont
  {Rojas}}, \bibinfo {author} {\bibfnamefont {S.}~\bibnamefont {Simpson}},
  \bibinfo {author} {\bibfnamefont {X.}~\bibnamefont {Chen}}, \bibinfo {author}
  {\bibfnamefont {D.~A.}\ \bibnamefont {Kunkel}}, \bibinfo {author}
  {\bibfnamefont {J.}~\bibnamefont {Nitz}}, \bibinfo {author} {\bibfnamefont
  {J.}~\bibnamefont {Xiao}}, \bibinfo {author} {\bibfnamefont {P.~A.}\
  \bibnamefont {Dowben}}, \bibinfo {author} {\bibfnamefont {E.}~\bibnamefont
  {Zurek}}, \ and\ \bibinfo {author} {\bibfnamefont {A.}~\bibnamefont
  {Enders}},\ }\href {\doibase 10.1039/c2cp40254h} {\bibfield  {journal}
  {\bibinfo  {journal} {Phys. Chem. Chem. Phys.}\ }\textbf {\bibinfo {volume}
  {14}},\ \bibinfo {pages} {4971} (\bibinfo {year} {2012})}\BibitemShut
  {NoStop}%
\bibitem [{\citenamefont {Wiberg}\ and\ \citenamefont
  {Rablen}(1993)}]{Wiberg1993}%
  \BibitemOpen
  \bibfield  {author} {\bibinfo {author} {\bibfnamefont {K.~B.}\ \bibnamefont
  {Wiberg}}\ and\ \bibinfo {author} {\bibfnamefont {P.~R.}\ \bibnamefont
  {Rablen}},\ }\href {\doibase 10.1002/jcc.540141213} {\bibfield  {journal}
  {\bibinfo  {journal} {J. Comput. Chem.}\ }\textbf {\bibinfo {volume} {14}},\
  \bibinfo {pages} {1504} (\bibinfo {year} {1993})}\BibitemShut {NoStop}%
\bibitem [{\citenamefont {Leenaerts}, \citenamefont {Partoens},\ and\
  \citenamefont {Peeters}(2008)}]{Leenaerts2008}%
  \BibitemOpen
  \bibfield  {author} {\bibinfo {author} {\bibfnamefont {O.}~\bibnamefont
  {Leenaerts}}, \bibinfo {author} {\bibfnamefont {B.}~\bibnamefont {Partoens}},
  \ and\ \bibinfo {author} {\bibfnamefont {F.~M.}\ \bibnamefont {Peeters}},\
  }\href {\doibase 10.1063/1.2949753} {\bibfield  {journal} {\bibinfo
  {journal} {Appl. Phys. Lett.}\ }\textbf {\bibinfo {volume} {92}},\ \bibinfo
  {pages} {243125} (\bibinfo {year} {2008})}\BibitemShut {NoStop}%
\bibitem [{\citenamefont {Jacquemin}\ \emph {et~al.}(2012)\citenamefont
  {Jacquemin}, \citenamefont {Bahers}, \citenamefont {Adamo}, \citenamefont
  {Ciofini},\ and\ \citenamefont {{Le Bahers}}}]{Jacquemin2012}%
  \BibitemOpen
  \bibfield  {author} {\bibinfo {author} {\bibfnamefont {D.}~\bibnamefont
  {Jacquemin}}, \bibinfo {author} {\bibfnamefont {T.~L.}\ \bibnamefont
  {Bahers}}, \bibinfo {author} {\bibfnamefont {C.}~\bibnamefont {Adamo}},
  \bibinfo {author} {\bibfnamefont {I.}~\bibnamefont {Ciofini}}, \ and\
  \bibinfo {author} {\bibfnamefont {T.}~\bibnamefont {{Le Bahers}}},\ }\href
  {\doibase 10.1039/C2CP40261K} {\bibfield  {journal} {\bibinfo  {journal}
  {Phys. Chem. Chem. Phys.}\ }\textbf {\bibinfo {volume} {14}},\ \bibinfo
  {pages} {5383} (\bibinfo {year} {2012})}\BibitemShut {NoStop}%
\bibitem [{\citenamefont {Hieringer}\ \emph {et~al.}(2011)\citenamefont
  {Hieringer}, \citenamefont {Flechtner}, \citenamefont {Kretschmann},
  \citenamefont {Seufert}, \citenamefont {Auw\"{a}rter}, \citenamefont {Barth},
  \citenamefont {G\"{o}rling}, \citenamefont {Steinr\"{u}ck},\ and\
  \citenamefont {Gottfried}}]{Hieringer2011}%
  \BibitemOpen
  \bibfield  {author} {\bibinfo {author} {\bibfnamefont {W.}~\bibnamefont
  {Hieringer}}, \bibinfo {author} {\bibfnamefont {K.}~\bibnamefont
  {Flechtner}}, \bibinfo {author} {\bibfnamefont {A.}~\bibnamefont
  {Kretschmann}}, \bibinfo {author} {\bibfnamefont {K.}~\bibnamefont
  {Seufert}}, \bibinfo {author} {\bibfnamefont {W.}~\bibnamefont
  {Auw\"{a}rter}}, \bibinfo {author} {\bibfnamefont {J.~V.}\ \bibnamefont
  {Barth}}, \bibinfo {author} {\bibfnamefont {A.}~\bibnamefont {G\"{o}rling}},
  \bibinfo {author} {\bibfnamefont {H.-P.}\ \bibnamefont {Steinr\"{u}ck}}, \
  and\ \bibinfo {author} {\bibfnamefont {J.~M.}\ \bibnamefont {Gottfried}},\
  }\href {\doibase 10.1021/ja1093502} {\bibfield  {journal} {\bibinfo
  {journal} {J. Am. Chem. Soc.}\ }\textbf {\bibinfo {volume} {133}},\ \bibinfo
  {pages} {6206} (\bibinfo {year} {2011})}\BibitemShut {NoStop}%
\bibitem [{\citenamefont {Boukari}, \citenamefont {Sonnet},\ and\ \citenamefont
  {Duverger}(2012)}]{Boukari2012}%
  \BibitemOpen
  \bibfield  {author} {\bibinfo {author} {\bibfnamefont {K.}~\bibnamefont
  {Boukari}}, \bibinfo {author} {\bibfnamefont {P.}~\bibnamefont {Sonnet}}, \
  and\ \bibinfo {author} {\bibfnamefont {E.}~\bibnamefont {Duverger}},\ }\href
  {\doibase 10.1002/cphc.201200578} {\bibfield  {journal} {\bibinfo  {journal}
  {ChemPhysChem}\ }\textbf {\bibinfo {volume} {13}},\ \bibinfo {pages} {3945}
  (\bibinfo {year} {2012})}\BibitemShut {NoStop}%
\bibitem [{Sup()}]{SuppInf}%
  \BibitemOpen
  \href@noop {} {}\bibinfo {note} {{See supplementary material at [url] for
  Figures S1-S5 and Tables SI-SVI.}}\BibitemShut {Stop}%
\bibitem [{\citenamefont {Bagus}, \citenamefont {Staemmler},\ and\
  \citenamefont {W\"{o}ll}(2002)}]{Bagus2002}%
  \BibitemOpen
  \bibfield  {author} {\bibinfo {author} {\bibfnamefont {P.~S.}\ \bibnamefont
  {Bagus}}, \bibinfo {author} {\bibfnamefont {V.}~\bibnamefont {Staemmler}}, \
  and\ \bibinfo {author} {\bibfnamefont {C.}~\bibnamefont {W\"{o}ll}},\ }\href
  {\doibase 10.1103/PhysRevLett.89.096104} {\bibfield  {journal} {\bibinfo
  {journal} {Phys. Rev. Lett.}\ }\textbf {\bibinfo {volume} {89}},\ \bibinfo
  {pages} {096104} (\bibinfo {year} {2002})}\BibitemShut {NoStop}%
\bibitem [{\citenamefont {Ferri}\ \emph {et~al.}(2015)\citenamefont {Ferri},
  \citenamefont {DiStasio}, \citenamefont {Ambrosetti}, \citenamefont {Car},\
  and\ \citenamefont {Tkatchenko}}]{Ferri2015}%
  \BibitemOpen
  \bibfield  {author} {\bibinfo {author} {\bibfnamefont {N.}~\bibnamefont
  {Ferri}}, \bibinfo {author} {\bibfnamefont {R.~A.}\ \bibnamefont {DiStasio}},
  \bibinfo {author} {\bibfnamefont {A.}~\bibnamefont {Ambrosetti}}, \bibinfo
  {author} {\bibfnamefont {R.}~\bibnamefont {Car}}, \ and\ \bibinfo {author}
  {\bibfnamefont {A.}~\bibnamefont {Tkatchenko}},\ }\href {\doibase
  10.1103/PhysRevLett.114.176802} {\bibfield  {journal} {\bibinfo  {journal}
  {Phys. Rev. Lett.}\ }\textbf {\bibinfo {volume} {114}},\ \bibinfo {pages}
  {176802} (\bibinfo {year} {2015})}\BibitemShut {NoStop}%
\bibitem [{\citenamefont {Mercurio}\ \emph {et~al.}(2013)\citenamefont
  {Mercurio}, \citenamefont {Maurer}, \citenamefont {Liu}, \citenamefont
  {Hagen}, \citenamefont {Leyssner}, \citenamefont {Tegeder}, \citenamefont
  {Meyer}, \citenamefont {Tkatchenko}, \citenamefont {Soubatch}, \citenamefont
  {Reuter},\ and\ \citenamefont {Tautz}}]{Mercurio2013}%
  \BibitemOpen
  \bibfield  {author} {\bibinfo {author} {\bibfnamefont {G.}~\bibnamefont
  {Mercurio}}, \bibinfo {author} {\bibfnamefont {R.~J.}\ \bibnamefont
  {Maurer}}, \bibinfo {author} {\bibfnamefont {W.}~\bibnamefont {Liu}},
  \bibinfo {author} {\bibfnamefont {S.}~\bibnamefont {Hagen}}, \bibinfo
  {author} {\bibfnamefont {F.}~\bibnamefont {Leyssner}}, \bibinfo {author}
  {\bibfnamefont {P.}~\bibnamefont {Tegeder}}, \bibinfo {author} {\bibfnamefont
  {J.}~\bibnamefont {Meyer}}, \bibinfo {author} {\bibfnamefont
  {A.}~\bibnamefont {Tkatchenko}}, \bibinfo {author} {\bibfnamefont
  {S.}~\bibnamefont {Soubatch}}, \bibinfo {author} {\bibfnamefont
  {K.}~\bibnamefont {Reuter}}, \ and\ \bibinfo {author} {\bibfnamefont {F.~S.}\
  \bibnamefont {Tautz}},\ }\href {\doibase 10.1103/PhysRevB.88.035421}
  {\bibfield  {journal} {\bibinfo  {journal} {Phys. Rev. B}\ }\textbf {\bibinfo
  {volume} {88}},\ \bibinfo {pages} {35421} (\bibinfo {year}
  {2013})}\BibitemShut {NoStop}%
\bibitem [{\citenamefont {Maurer}, \citenamefont {Ruiz},\ and\ \citenamefont
  {Tkatchenko}(2015)}]{Maurer2015}%
  \BibitemOpen
  \bibfield  {author} {\bibinfo {author} {\bibfnamefont {R.~J.}\ \bibnamefont
  {Maurer}}, \bibinfo {author} {\bibfnamefont {V.~G.}\ \bibnamefont {Ruiz}}, \
  and\ \bibinfo {author} {\bibfnamefont {A.}~\bibnamefont {Tkatchenko}},\
  }\href {\doibase 10.1063/1.4922688} {\bibfield  {journal} {\bibinfo
  {journal} {J. Chem. Phys.}\ }\textbf {\bibinfo {volume} {143}},\ \bibinfo
  {pages} {102808} (\bibinfo {year} {2015})}\BibitemShut {NoStop}%
\bibitem [{\citenamefont {Tkatchenko}\ \emph {et~al.}(2012)\citenamefont
  {Tkatchenko}, \citenamefont {DiStasio}, \citenamefont {Car},\ and\
  \citenamefont {Scheffler}}]{Tkatchenko2012}%
  \BibitemOpen
  \bibfield  {author} {\bibinfo {author} {\bibfnamefont {A.}~\bibnamefont
  {Tkatchenko}}, \bibinfo {author} {\bibfnamefont {R.~A.}\ \bibnamefont
  {DiStasio}}, \bibinfo {author} {\bibfnamefont {R.}~\bibnamefont {Car}}, \
  and\ \bibinfo {author} {\bibfnamefont {M.}~\bibnamefont {Scheffler}},\ }\href
  {\doibase 10.1103/PhysRevLett.108.236402} {\bibfield  {journal} {\bibinfo
  {journal} {Phys. Rev. Lett.}\ }\textbf {\bibinfo {volume} {108}},\ \bibinfo
  {pages} {236402} (\bibinfo {year} {2012})}\BibitemShut {NoStop}%
\bibitem [{\citenamefont {Joshi}\ \emph {et~al.}(2014)\citenamefont {Joshi},
  \citenamefont {Bischoff}, \citenamefont {Koitz}, \citenamefont {Ecija},
  \citenamefont {Seufert}, \citenamefont {Seitsonen}, \citenamefont {Hutter},
  \citenamefont {Diller}, \citenamefont {Urgel}, \citenamefont {Sachdev},
  \citenamefont {Barth},\ and\ \citenamefont {Auw\"{a}rter}}]{Joshi2014}%
  \BibitemOpen
  \bibfield  {author} {\bibinfo {author} {\bibfnamefont {S.}~\bibnamefont
  {Joshi}}, \bibinfo {author} {\bibfnamefont {F.}~\bibnamefont {Bischoff}},
  \bibinfo {author} {\bibfnamefont {R.}~\bibnamefont {Koitz}}, \bibinfo
  {author} {\bibfnamefont {D.}~\bibnamefont {Ecija}}, \bibinfo {author}
  {\bibfnamefont {K.}~\bibnamefont {Seufert}}, \bibinfo {author} {\bibfnamefont
  {A.~P.}\ \bibnamefont {Seitsonen}}, \bibinfo {author} {\bibfnamefont
  {J.}~\bibnamefont {Hutter}}, \bibinfo {author} {\bibfnamefont
  {K.}~\bibnamefont {Diller}}, \bibinfo {author} {\bibfnamefont {J.~I.}\
  \bibnamefont {Urgel}}, \bibinfo {author} {\bibfnamefont {H.}~\bibnamefont
  {Sachdev}}, \bibinfo {author} {\bibfnamefont {J.~V.}\ \bibnamefont {Barth}},
  \ and\ \bibinfo {author} {\bibfnamefont {W.}~\bibnamefont {Auw\"{a}rter}},\
  }\href {http://dx.doi.org/10.1021/nn406024m} {\bibfield  {journal} {\bibinfo
  {journal} {ACS Nano}\ }\textbf {\bibinfo {volume} {8}},\ \bibinfo {pages}
  {430} (\bibinfo {year} {2014})}\BibitemShut {NoStop}%
\end{thebibliography}
\end{document}